\documentclass[fleqn,usenatbib]{mnras}

\usepackage{times}

\usepackage[T1]{fontenc}
\usepackage{ae,aecompl}

\usepackage{graphicx}   
\usepackage{amsmath}    
\usepackage{amssymb}    
\usepackage{hyperref}
\usepackage{booktabs,tabularx}
\usepackage{threeparttable}
\usepackage{enumitem}
\usepackage{textcomp}
\usepackage{gensymb}
\usepackage{makecell}
\usepackage{bm}

\usepackage{pgf}
\usepackage{soul}
\definecolor{lightgreen}{HTML}{B7F774}
\definecolor{lightred}{HTML}{FF6666}
\definecolor{lightorange}{HTML}{FE9A2E}
\sethlcolor{lightgreen}

\newcommand{\fms}[1]{}
\newcommand{\lvek}[1]{}
\newcommand{\lvekhidden}[1]{}
\newcommand{\refs}[1]{\textcolor{blue}{\,\bf [References]}\,}

\title[LOFAR-EoR 21-cm power spectrum upper limit]{Improved upper limits on the 21-cm signal power spectrum of neutral hydrogen at $\bm{z \approx 9.1}$ from LOFAR}

\author[F. G. Mertens et al.]{F. G. Mertens$^{1,3}$\thanks{E-mail: mertens@astro.rug.nl},
M. Mevius$^{2}$\thanks{E-mail: mevius@astron.nl},
L.V.E Koopmans$^{1}$,
A. R. Offringa$^{2}$,
G. Mellema$^{4}$,
\newauthor
S. Zaroubi$^{5,6,1}$,
M. A. Brentjens$^{2}$,
H. Gan$^{1}$,
B. K. Gehlot$^{7}$,
V. N. Pandey$^{2}$,
\newauthor
A. M. Sardarabadi$^{1}$,
H. K. Vedantham$^{2}$, 
S. Yatawatta$^{2}$,
K. M. B. Asad$^{8}$,
B. Ciardi$^{9}$,
\newauthor
E. Chapman$^{10}$,
S. Gazagnes$^{1}$,
R. Ghara$^{4,5,6}$, 
A. Ghosh$^{11,12,13}$,
S. K. Giri$^{4}$,
I. T. Iliev$^{14}$,
\newauthor
V. Jeli\'c$^{15}$,
R. Kooistra$^{16}$,
R. Mondal$^{14}$,
J. Schaye$^{17}$,
M. B. Silva$^{18}$
\\
\\
$^{1}$Kapteyn Astronomical Institute, University of Groningen, PO Box 800, 9700 AV Groningen, The Netherlands \\
$^{2}$Astron, PO Box 2, 7990 AA Dwingeloo, The Netherlands \\
$^{3}$LERMA, Observatoire de Paris, PSL Research University, CNRS, Sorbonne Universit\'e, F-75014 Paris, France
\\
$^{4}$The Oskar Klein Centre, Department of Astronomy, Stockholm University, AlbaNova, SE-10691 Stockholm, Sweden\\
$^{5}$Department of Natural Sciences, The Open University of Israel, 1 University Road, PO Box 808, Ra'anana 4353701, Israel \\
$^{6}$Department of Physics, Technion, Haifa 32000, Israel\\
$^{7}$School of Earth and Space Exploration, Arizona State University, 781 Terrace Mall, Tempe, AZ 85287, U.S.A.\\
$^{8}$Independent University Bangladesh, Plot 16, Block B, Aftabuddin Ahmed Road, Bashundhara R/A, Dhaka, Bangladesh \\
$^{9}$Max-Planck Institute for Astrophysics, Karl-Schwarzschild-Stra{\ss}e 1, 85748 Garching, Germany\\
$^{10}$Astrophysics Group, Imperial College London, Blackett Laboratory, Prince Consort Road, London, SW7 2AZ, United Kingdom\\
$^{11}$Department of Physics, University of the Western Cape, Cape Town 7535, South Africa \\
$^{12}$SARAO, 2 Fir Street, Black River Park, Observatory, Capetown, South Africa\\
$^{13}$Department of Physics, Banwarilal Bhalotia College, Asansol, West Bengal, India\\
$^{14}$Astronomy Centre, Department of Physics and Astronomy, Pevensey II Building, University of Sussex, Brighton BN1 9QH, U.K.\\
$^{15}$Ru{\dj}er Bo\v{s}kovi\'{c} Institute, Bijeni\v{c}ka cesta 54, 10000 Zagreb, Croatia\\
$^{16}$Kavli IPMU (WPI), UTIAS, The University of Tokyo, Kashiwa, Chiba 277-8583, Japan\\
$^{17}$Leiden Observatory, Leiden University, PO Box 9513, 2300RA Leiden, The Netherlands\\
$^{18}$Institute of Theoretical Astrophysics, University of Oslo, PO Box 1029 Blindern, N-0315 Oslo, Norway\\
\vspace{-1em}
}
\date{Accepted XXX. Received YYY; in original form ZZZ}

\pubyear{2019}

\begin{document}
\label{firstpage}
\pagerange{\pageref{firstpage}--\pageref{lastpage}}
\maketitle

\begin{abstract}
A new upper limit on the 21-cm signal power spectrum at a redshift of $z \approx 9.1$ is presented, based on 141 hours of data obtained with the Low-Frequency Array (LOFAR). The analysis includes significant improvements in spectrally-smooth gain-calibration, Gaussian Process Regression (GPR) foreground mitigation and optimally-weighted power spectrum inference. Previously seen `excess power' due to spectral structure in the gain solutions has markedly reduced but some excess power still remains with a spectral correlation distinct from thermal noise. This excess has a spectral coherence scale of $0.25 - 0.45$\,MHz and is partially correlated between nights, especially in the foreground wedge region. The correlation is stronger between nights covering similar local sidereal times. A best 2-$\sigma$ upper limit of $\Delta^2_{21} < (73)^2\,\mathrm{mK^2}$ at $k = 0.075\,\mathrm{h\,cMpc^{-1}}$ is found, an improvement by a factor $\approx 8$ in power compared to the previously reported upper limit. The remaining excess power could be due to residual foreground emission from sources or diffuse emission far away from the phase centre, polarization leakage, chromatic calibration errors, ionosphere, or low-level radio-frequency interference. We discuss future improvements to the signal processing chain that can further reduce or even eliminate these causes of excess power. 
\end{abstract}




\section{Introduction}
\label{sec:intro}
Exploring the Cosmic Dawn (CD) and the subsequent Epoch of Reionization (EoR), comprising two eras from $z\sim 6-30$ when the first stars, galaxies and black holes heated and ionized the Universe, is of great importance to our understanding of the nature of these first radiating sources. It provides insight on the timing and mechanisms of their formation, as well as the impact on the physics of the interstellar medium (ISM) and intergalactic medium (IGM) of the radiation emitted by these first light sources~\citep[see, e.g.][for extensive reviews]{Ciardi05,Morales10,Pritchard12,Furlanetto16}.

Observations of the Gunn-Peterson trough in high-redshift quasar spectra~\citep[e.g.][]{Becker01, Fan06} and the measurement of the optical depth to Thomson scattering of the Cosmic Microwave Background (CMB) radiation~\citep[e.g.][]{Planck16} both suggest that the bulk of reionization took place in the redshift range $6 \lesssim z \lesssim 10$. The evolution of the observed Ly-$\alpha$ Emitter (LAE) luminosity function at $z > 6$~\citep{Clement12, Schenker13} and the Ly-$\alpha$ absorption profile toward very distant quasars~\citep{Mortlock16,Greig17,Davies18} are other indirect probes of the EoR. 

\lvek{Update to recent high-z quasar at z=7.5} \fms{I added references for recent work on the z=7.5 quasar. Do we also want to quote some estimate of neutral fraction here ?}

The most direct probe of this epoch, however, is the redshifted 21-cm line from neutral hydrogen, seen in emission or absorption against the CMB~\citep{Madau97,Shaver99,Tozzi00,zaroubi13}. A number of observational programs are currently underway, or have recently been completed that aimed to detect the 21-cm brightness temperature from the EoR and CD. The 21-cm global experiments, such as EDGES\footnote{Experiment to Detect the Global Epoch of Reionization Signature, https://loco.lab.asu.edu/edges/}~\citep{Bowman18} or SARAS\footnote{Shaped Antenna measurement of the background RAdio Spectrum, http://www.rri.res.in/DISTORTION/saras.html}~\citep{Singh17} aim to measure the sky-averaged spectrum of the 21-cm signal. The tentative detection of the global 21-cm signal reported by the EDGES team~\citep{Bowman18} has unexpected properties. This signal, consisting of a flat-bottomed deep absorption-line feature during the CD at $z = 14 - 21$, is considerably stronger and wider than predicted~\citep{Fraser18}, and, depending on the additional mechanism invoked to explain it~\citep[e.g.][]{Barkana18b,Berlin18,Ewall18,Fialkov19,Mirocha19}, could also have an impact on the predicted strength of the 21-cm brightness temperature fluctuations during the EoR. Complementary to these, the interferometric experiments aim at a statistical detection of the  fluctuations from the EoR using radio interferometers such as LOFAR\footnote{Low-Frequency Array, http://www.lofar.org}, MWA\footnote{Murchison Widefield Array, http://www.mwatelescope.org} or PAPER\footnote{Precision Array to Probe EoR, http://eor.berkeley.edu}. 

These instruments have already set impressive upper limits on the 21-cm signal power spectra, considering the extreme challenges they face, but have not yet achieved a detection. Using the GMRT\footnote{Giant Metrewave Radio Telescope, http://gmrt.ncra.tifr.res.in},~\cite{Paciga13} reported a $2-\sigma$ upper limit of $\Delta_{21}^2 < (248\, \mathrm{mK})^2$ at $z = 8.6$ and wave-number $k \approx 0.5 \mathrm{h\,cMpc^{-1}}$ from a total of about 40 hours of observed data. Recently,~\cite{Barry19} reported a $2-\sigma$ upper limit of $\Delta_{21}^2 < (62.4 \mathrm{mK})^2$ at $z = 7$ and $k \approx 0.2 \mathrm{h\,cMpc^{-1}}$ using 21 hours of Phase I MWA data, and~\cite{Li19} published  a $2-\sigma$ upper limit of $\Delta_{21}^2 < (49 \mathrm{mK})^2$  at $z = 6.5$ and $k \approx 0.59 \mathrm{h\,cMpc^{-1}}$ using 40 hours of Phase II MWA data. The PAPER collaboration reported a very deep upper limit~\citep{Ali15}, but after re-analysis~\citep{Cheng18} have recently reported revised and higher upper limits~\citep{Kolopanis19}, the deepest being $\Delta_{21}^2 < (200 \mathrm{mK})^2$ at $z = 8.37$ and $k \approx 0.37 \mathrm{h\,cMpc^{-1}}$. In~\cite{Patil17}, the LOFAR-EoR Key Science Project (KSP) published their first upper limit based on 13h of data from LOFAR, reporting a $2-\sigma$ upper limit of $\Delta_{21}^2 < (79.6 \mathrm{mK})^2$ at $z = 10.1$ and $k \approx 0.053 \mathrm{h\,cMpc^{-1}}$.

\lvek{Why these refs?} \fms{I changed the ref to `historical' high-z 21-cm work} 

Much more research is still needed, however, to control the many complex aspects in the signal processing chain~\citep{Liu19} in order to reach the expected 21-cm signal strengths which lie two to three orders of magnitude below these limits~\citep[e.g.][]{Mesinger11}. Mitigating all possible effects that could prevent a 21-cm signal detection is particularly important since these instruments are also pathfinders for the much more sensitive and ambitious second-generation instruments such as the SKA\footnote{Square Kilometre Array, http://www.skatelescope.org}~\citep{Koopmans15} and HERA\footnote{Hydrogen Epoch of Reionization Array, http://reionization.org}~\citep{DeBoer17}. 

At the low radio-frequencies targeted by 21-cm signal observations, the radiation from the Milky Way and other extragalactic sources dominates the sky by many orders of magnitude in brightness~\citep{Shaver99}. The emission of these foregrounds varies smoothly with frequency, and this characteristic can be used to differentiate it from the rapidly fluctuating 21-cm signal~\citep{Jelic08}. However, due to the ionosphere and the frequency-dependent response of the radio telescopes (e.g. its primary beam and $uv$-coverage both scale with frequency),  structure is introduced to the otherwise spectrally-smooth foregrounds, causing so-called `mode-mixing'~\citep{Morales12}. Most of these chromatic effects are confined inside a wedge-like shape in $k$-space~\citep{Datta10, Trott12, Vedantham12, Liu14a, Liu14b}, and to mitigate them, many experiments adopt a  `foreground avoidance' strategy which only performs statistical analyses of the 21-cm signal inside a region in $k$-space where the thermal noise and 21-cm signals dominate~\citep[e.g.][]{Jacobs16,Kolopanis19}. In practice, however, leakage above the wedge is also observed and is thought to be due to gain-calibration errors because of an incomplete or incorrect sky model~\citep{Patil16,Ewall17}, errors in band-pass calibration, cable reflections~\citep{Beardsley16}, multi-path propagation, mutual coupling~\citep{kern19a}, residual radio-frequency interference (RFI)~\citep{Whitler19,Offringa19a}, as well as chromatic errors introduced due to leakage from the polarized sky into Stokes I~\citep{Jelic10,Spinelli18} or ionospheric disturbances~\citep{Koopmans10,Vedantham16}.

By modelling and removing the foreground contaminants, the LOFAR EoR KSP team aims at probing the 21-cm signal both outside and inside the wedge, thereby potentially increasing the sensitivity to the 21-cm signal by an order of magnitude~\citep{Pober14} and enabling exploration of the signal at the largest available scales, which have more significance for cosmology/signal-clustering studies. This has required the development of a comprehensive sky model of the North Celestial Pole (NCP) field~\citep{Yatawatta13, Patil17}, currently consisting of nearly thirty thousand components. The model is used to solve station gains in a large number of directions using the distributed gain-calibration code \texttt{Sagecal-CO}\footnote{https://github.com/nlesc-dirac/sagecal}~\citep{Yatawatta16}, and subsequently removes these components with their direction-dependent instrumental response functions. Confusion-limited residual compact and diffuse foregrounds also need to be removed and, to this end, we employ a novel strategy consisting of statistically separating the contribution of the 21-cm signal from the foregrounds using the technique of Gaussian Process Regression~\citep[GPR,][]{Mertens18, Gehlot19a}. These data processing steps are described in Section~\ref{sec:formalism}.

We report here an improved 21-cm power spectrum upper limit from the LOFAR EoR Key Science Project based on a total of ten nights of observations (141 hours of data) of the NCP field, acquired during the first three LOFAR cycles. In this work, we focus on the redshift bin $z \approx 8.7 - 9.6$, corresponding to the frequency range $134 - 146$ MHz. Our observational strategy is described in Section~\ref{sec:observation}. The processing and analyses of these observations are discussed in Sections~\ref{sec:formalism} and~\ref{sec:results_individual}. A new upper limit on the 21-cm signal power spectra is presented in Section~\ref{sec:results_combined}. Finally, we discuss the remaining excess power (in comparison with the thermal noise power) that we observe, its potential origins, and improvements of the processing pipeline that we aim to implement to reduce it, in Section~\ref{sec:discussion}. The implications of this improved upper limit are studied in~\cite{Ghara20} and a summary of their finding is also presented in Section~\ref{sec:implication}. Throughout this paper we use a $\Lambda$CDM cosmology consistent with the Planck 2015 results~\citep{Planck16XIII}. All distances and wavenumbers are in comoving coordinates.

\begin{table*}
\centering
\caption{List of all the nights of observation analysed in this work.
Information on observation date, time and duration, along with noise statistics
is given for every nights.}
\begin{threeparttable}
\label{tab:all_nights}
\begin{tabular}{lrrrrrrr}
\toprule
\makecell{Night ID\\\,} & \makecell{LOFAR\\Cycle} & \makecell{UTC observing start\\date and time} & \makecell{LST$^a$ starting\\time [hour]}& \makecell{Duration [hour]\\\,} & \makecell{SEFD$^b$ estimate \\\,[Jy]} & \makecell{$ \frac{<|\delta_\nu V_V|^2>} {<|\delta_t V_I|^2>}^c$ } & \makecell{$\frac{<|\delta_\nu V_I|^2>} {<|\delta_t V_I|^2>}^d$} \\
\midrule
L80847 & 0 & 2012-12-31 15:33:06 & 22.7 & 16.0 & 4304 & 1.28 & 1.88 \\
L80850$^*$ & 0 & 2012-12-24 15:30:06 & 22.2 & 16.0 & 4226 & 1.61 & 2.19 \\
L86762 & 0 & 2013-02-06 17:20:06 & 2.9 & 13.0 & 4264 & 1.30 & 1.93 \\
L90490 & 0 & 2013-02-11 17:20:06 & 3.2 & 13.0 & 4331 & 1.32 & 1.91 \\
L196421 & 1 & 2013-12-27 15:48:38 & 22.7 & 15.5 & 4077 & 1.62 & 2.21 \\
L205861 & 1 & 2014-03-06 17:46:30 & 5.2 & 11.9 & 3884 & 1.37 & 1.92 \\
L246297 & 2 & 2014-10-23 16:46:30 & 19.3 & 13.0 & 4294 & 1.31 & 1.95 \\
L246309 & 2 & 2014-10-16 17:01:41 & 19.1 & 12.6 & 4253 & 1.24 & 1.60 \\
L253987 & 2 & 2014-12-05 15:44:35 & 21.1 & 15.3 & 3978 & 1.23 & 1.88 \\
L254116 & 2 & 2014-12-10 15:42:54 & 21.4 & 15.4 & 4298 & 1.21 & 1.80 \\
L254865 & 2 & 2014-12-23 15:45:36 & 22.3 & 15.5 & 4057 & 1.31 & 1.88 \\
L254871$^*$ & 2 & 2014-12-20 15:44:04 & 22.1 & 15.5 & 3917 & 1.25 & 1.73 \\
\bottomrule
\end{tabular}
\begin{tablenotes}
  \small
  \item[a] Local Sidereal Time.
  \item[b] System Equivalent Flux Density.
  \item[c] Ratio of Stokes V sub-band difference power over thermal
  noise power.
  \item[d] Ratio of Stokes I sub-band difference power over thermal
  noise power.
  \item[*] These two nights are not part of the 10 nights selection.
\end{tablenotes}
\end{threeparttable}
\end{table*}

\section{LOFAR-HBA Observations}
\label{sec:observation}

The LOFAR EoR KSP targets mainly two deep fields: the NCP and the field surrounding the bright compact radio source 3C\,196~\citep{deBruyn12}. Here we present results on the NCP field for which we already published an upper limit on the 21-cm signal based on 13 hours of data~\citep{Patil17}. The NCP can be observed every night of the year, making it an excellent EoR window. Currently $\approx 2480$ hours of data have been observed with the LOFAR High-Band Antenna (HBA) system. The LOFAR HBA radio interferometer consists of 24 core stations distributed over an area of about 2\,km diameter, 14 remote stations distributed over the Netherlands, providing a maximum baseline length of $\sim 100$\,km, and an increasing number of international stations distributed over Europe~\citep{vanHaarlem13}. In this work, we analysed 12 nights of observations from the LOFAR Cycle 0, 1 and 2. The observations are carried out using all core stations (in split mode, so de facto providing 48 stations) and remote stations\footnote{The remote stations, which comprise nominally 48 tiles compared to the 24 tiles of a split core station, were tapered to have the same size and shape as the core stations.} in the frequency range from 115 to 189 MHz, with a spectral resolution of $3.05\,$kHz (i.e. 64 channels per sub-band of $195.3\,$kHz width), and a temporal resolution of 2 seconds. NCP observations were scheduled from `dusk to dawn' (thus avoiding strong ionospheric effects and avoiding the sun), and have a typical duration of $12 - 16$ h. While data have been acquired over the $115 - 189$ MHz band, we concentrate our effort in the current work on the redshift bin $z \approx 8.7 - 9.6$ (frequency range $134 - 146$ MHz), thus reducing the required processing time while we are further optimizing our calibration strategy. The observational details of the different nights analyzed are summarized in Table~\ref{tab:all_nights}. \lvek{Add a panel of images of the NCP from AARTFAAC(?) and LOFAR.} \fms{Figure added, images will be improved in later iteration.}

\section{Methodology and Data Processing}
\label{sec:formalism}

\begin{figure*}
\includegraphics{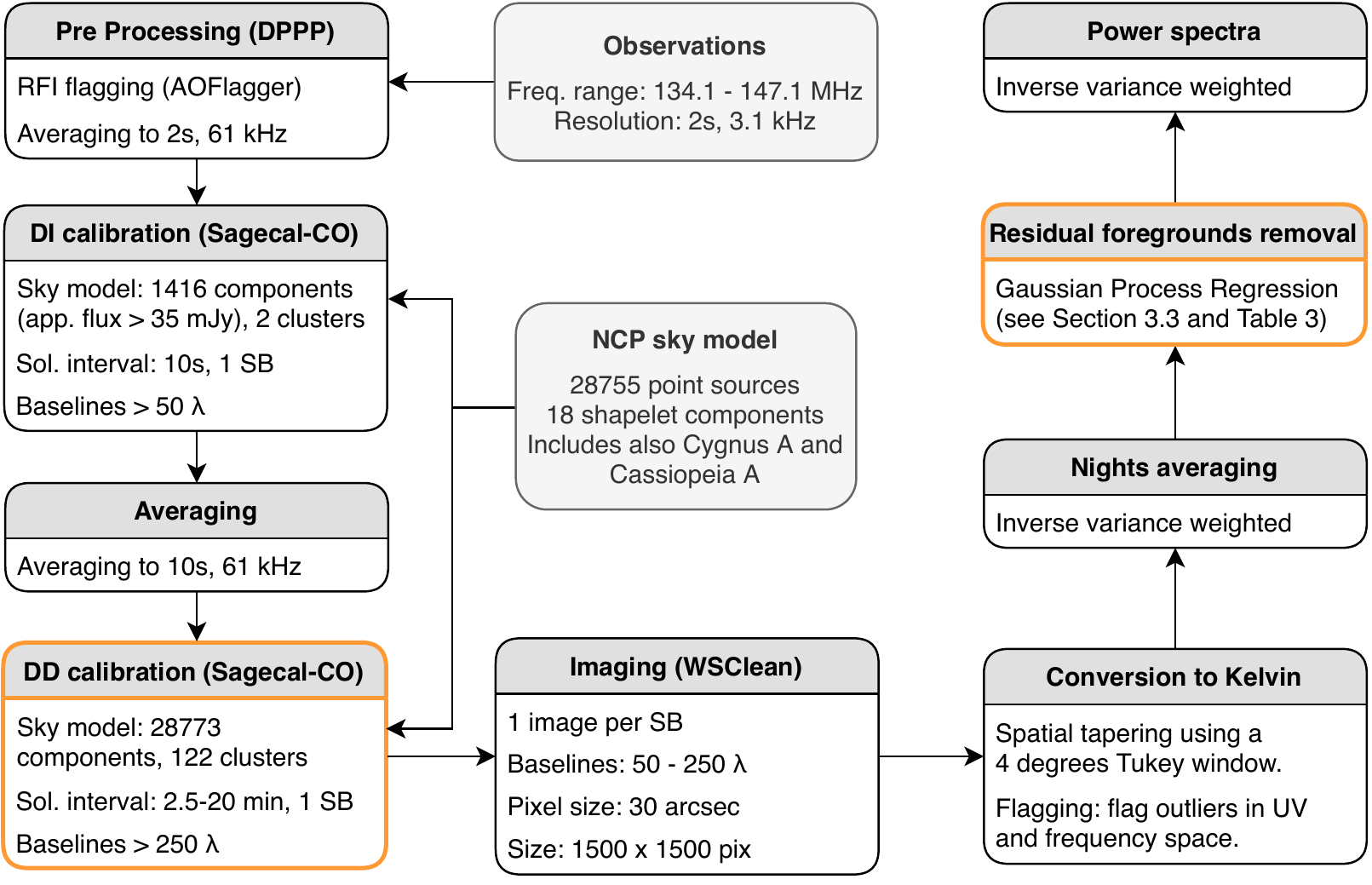}
\caption{\label{fig:lofar_hba_ncp_processing} The LOFAR-EoR HBA processing pipeline, describing the steps required to reduce the raw observed visibilities to the 21-cm signal power spectra. The development of the sky-model used at the calibration steps is not described here. The orange outline denotes processes of the pipeline which can have a substantial impact on the 21-cm signal and which are tested through signal injection and simulation~(see Section~\ref{sec:cross-check} and Mevius~et~al.~in prep.).}
\end{figure*}

We first introduce the methods and processing steps used to reduce the data from the raw observed visibilities to the power spectra. The LOFAR-EoR data processing pipeline consists, in essence, of (1) Pre-processing and RFI excision, (2) direction-independent calibration (DI-calibration), (3) direction-dependent calibration (DD-calibration) including subtraction of the sky-model, (4) imaging, (5) residual foregrounds modelling and removal, (6) power spectra estimation. The strategy used in steps (1) and (2) is similar to the one adopted in~\cite{Patil17} while the strategy used for the rest of the steps has undergone significant revisions. Figure~\ref{fig:lofar_hba_ncp_processing} shows an overview of the LOFAR-EoR data processing pipeline. All data processing is performed on a dedicated compute-cluster called Dawn~\citep{Pandey2020}, which consists of 48 $\times$ 32 hyperthreaded compute cores and 124 Nvidia K40 GPUs. The cluster is located at the Centre for Information Technology of the University of Groningen.

\subsection{Calibration and Imaging}
\label{sec:cal_img}

In this section, we describe the processes involved in transforming uncalibrated observed visibilities to calibrated, sky-model subtracted image cubes.

\subsubsection{RFI flagging}
\label{sec:rfi_flag}

RFI-flagging is done on the highest time and frequency resolution data (2 seconds, 64 channels per sub-band) using \texttt{AOflagger}\footnote{https://sourceforge.net/projects/aoflagger/}~\citep{Offringa12}. The four edge channels of the 64 sub-band channels, each having 3.05\,kHz spectral resolution, affected by aliasing from the poly-phase filter, are also flagged. This reduces the effective width of a sub-band to 183 kHz. The data is then averaged to 15 channels (12.2\,kHz) per sub-band to reduce the data volume for archiving purposes and further processing (all LOFAR-EoR observations are archived in the LOFAR LTA at surfSARA, and Poznan). It was later found that the data was not correctly flagged during this first RFI flagging stage (the time-window was of insufficient size to correctly detect time-correlated RFI). Since the highest resolution on which the data is archived is 15 channels per sub-band and 2 seconds, we decided to apply a second RFI flagging on these data before averaging to the three channels and 2 seconds data product which is used in the initial steps of the calibration. The intra-station baselines of length 127 m share the same electronics cabinet and are prone to correlated RFI generated inside the cabinet itself. Hence, these baselines are also flagged during the preprocessing step. Typically about 5\% of visibilities are flagged at this stage~\citep{Offringa13}.

\subsubsection{The NCP sky model}
\label{sec:sky_model}

The source model components of the NCP field~\citep{Bernardi10,Yatawatta13} has been iteratively built over many years from the highest resolution images, with an angular resolution $\approx$ 6 arcsec, using \texttt{buildsky}~\citep{Yatawatta13}. This sky model is composed of 28773 unpolarized components (28755 delta functions and 18 shaplets\footnote{Shapelets form an orthonormal basis in which a source of arbitrary shape can be described by a limited number of coefficients with sufficient accuracy~\protect\citep{Yatawatta11}.}) covering all sources up to 19 degrees distance from the NCP and down to an apparent flux density of $\approx 3\ \mathrm{mJy}$ inside the primary beam. It also includes Cygnus~A about $50\degree$ away from the NCP, and Cassiopeia~A about $30\degree$ away from the NCP, which are the two brightest radio sources in the Northern hemisphere. The spectra of each component are modeled by a third order polynomial function in log-log space. For modeling some of the brightest sources we have also made use of international baselines in LOFAR, which provide a resolution down to 0.25 arcsec. 

\begin{figure*}
    \includegraphics{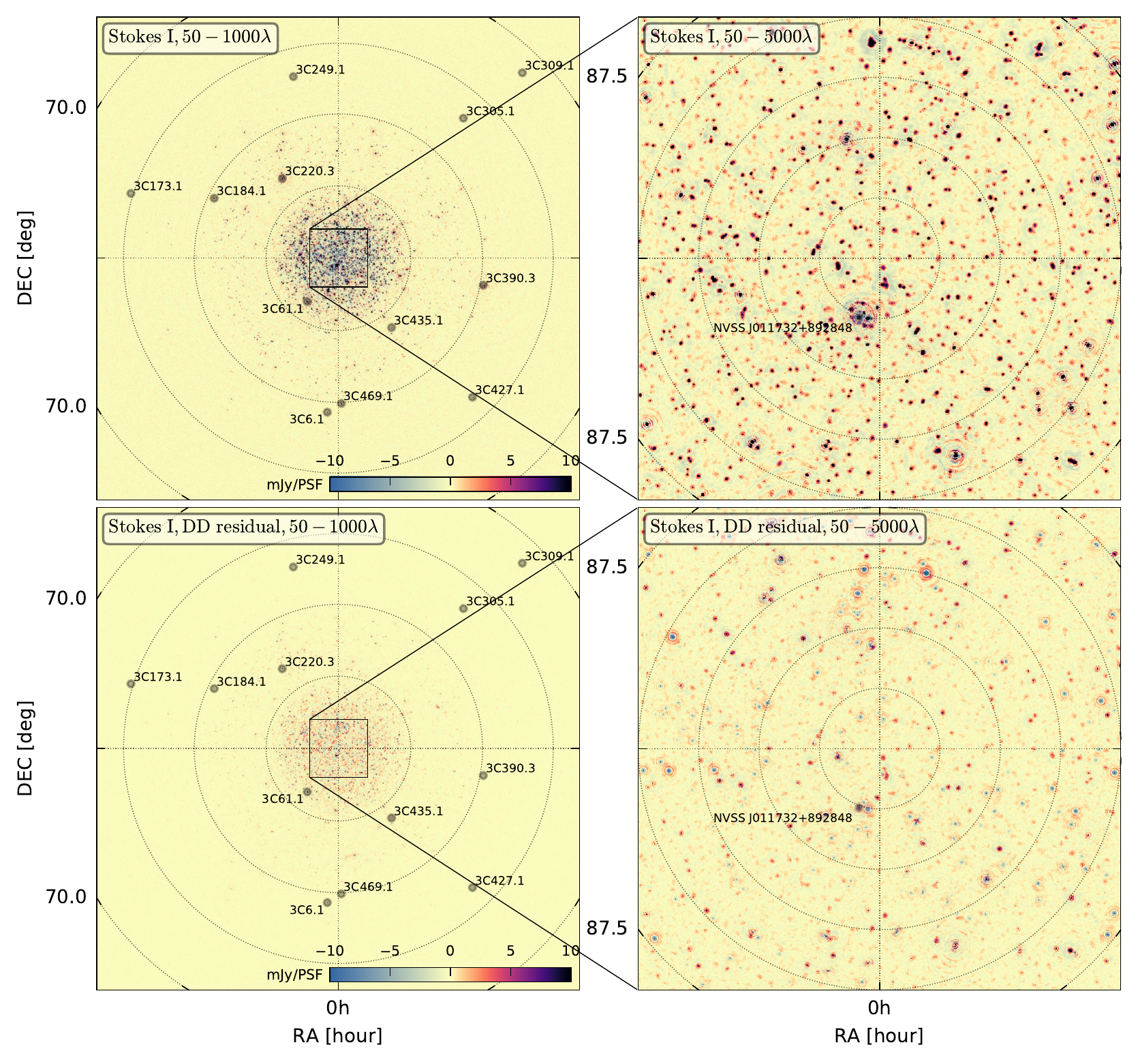}
    \caption{\label{fig:clean_img} LOFAR-HBA Stokes I continuum images ($134 - 146$ MHz) of the NCP field. All 12 nights ($\approx 170$ hours) were included in making these images. The top panels show the field after DI calibration, with 3C61.1 subtracted in the visibilities using \texttt{Sagecal}, and the images deconvolved using \texttt{WSClean}. The bottom panels show the residual after DD-calibration. The left panels show a $34\degree \times 34\degree$ image with a resolution of 3.5 arcmin (baselines between $50 - 1000 \lambda$) and include the positions of the 3C sources in the field (black circles). The right panels are zoomed $4\degree \times 4\degree$ images with a resolution of 42 arcsec (baselines between $50 - 5000 \lambda$) in which we also indicate the position of NVSS~J011732+892848 (black circle). Power spectra are measured in this $4\degree \times 4\degree$ field of view.}
\end{figure*}

The intensity scale of our sky model is set by NVSS~J011732+892848 (RA~01h\,17m\,33s, Dec~89$\degree$\,28'\,49'' in J2000) (see Fig.~\ref{fig:clean_img}), a flat spectrum source with an intrinsic flux of 8.1 Jy with 5\% accuracy~\citep{Patil17}. The flux and spectrum of this source were obtained following a calibration against 3C295 in the range $120 - 160$ MHz~\citep{Patil17}. Fig.~\ref{fig:clean_img} (top panels) shows images of the NCP field after DI calibration, revealing the sources with flux > 3 mJy in the inner $4\degree \times 4\degree$ and sources observable at a distance up to $15 \degree$ from the phase center (up to the second side-lobe of the LOFAR-HBA primary beam). Many of these sources have complex spatial structure and are modeled by multiple delta functions (or shaplets). The accuracy of our flux scale calibration is tested by cross-identifying the 100 brightest sources observed at a distance $< 3 \degree$ from the phase center with the 6C~\citep{Baldwin85} and 7C~\citep{Hales07} 151 MHz radio catalogs. We obtained the intrinsic flux of these sources by first applying a primary-beam correction, and then modeling their spectra over the 13 MHz bandwidth with a power-law to estimate their flux at 151 MHz. We found a mean ratio of 1.02 between our intrinsic flux and the 6C/7C flux with a standard deviation of 0.12, highlighting the accuracy of our absolute flux scale calibration. We additionally found that the night-to-night fluctuations of the flux of these bright sources are on average about 5\%, likely due to intrinsic sources fluctuations and primary beam errors not captured by the DI-calibration step.

\subsubsection{Direction-independent calibration}
\label{sec:di_cal}

For direction-independent calibration, we use the same approach as described in~\cite{Patil17}. Since the relatively bright source in the NCP field, 3C\,61.1 (see Fig.~\ref{fig:clean_img}), is close to the first null of the station's primary beam, it is necessary to have a separate set of solutions for this direction. In that way we isolate the strong direction-dependent effects of this source. The remainder of the field is modeled by selecting the 1416 brightest components from the NCP sky model, down to an apparent flux limit of 35\,mJy. This flux limit was chosen to reduce the processing time while still preserving the signal-to-noise (S/N) required to calibrate the instrument toward these two directions at high time resolution, the power of the remaining sources in the 28773 components NCP sky model account for only 1\% of the total power of the sky model. Calibration is performed on the three channels (61\,kHz), and 2 second resolution data set with a spectral and time solution interval of 195.3 kHz (one sub-band) and 10 seconds, thus allowing to solve for fast direction-independent ionospheric phase variations. Calibration is done using \texttt{Sagecal-CO}~\citep{Yatawatta16}, constraining the solutions in frequency with a third-order Bernstein polynomial over 13\,MHz bandwidth. \texttt{Sagecal}'s consensus optimization distributes the processing over several compute nodes while iteratively penalizing solutions that deviate from a frequency smooth prior by a quadratic regularization term. The frequency smooth prior is updated at each iteration. If given a sufficient number of iterations, this process should converge to this prior. We refer the readers to~\cite{Yatawatta15,Yatawatta16} for a more detailed description of the \texttt{Sagecal-CO} algorithm. In addition to smooth spectral gain variations, we also solve at this stage for the fast frequency varying band-pass response of the stations, which are caused by low-pass and high-pass filters in the signal chain as well as reflections in the coax-cables between tiles and receivers~\citep{Offringa13,Beardsley16,kern19b}. For this purpose, we use a low regularization parameter and limit the number of iterations to 20. After DI-calibration, outliers in the visibilities (with an amplitude conservatively set to be larger than 70\,Jy) are flagged and the data are averaged to the final data product of 3 channels and 10 seconds.

\lvekhidden{Confusing. Why 1300 here and 28000 above? Need to explain why fewer components are used an why. Speed? Also how are these 1300 source selected and down to what flux level?} \lvekhidden{Provide spectral and time resolution for these solutions.} \lvekhidden{Error on this flux? Is this the average flux over 12h or 16h? Since the beam changes, this might make a difference.} \lvekhidden{Explain "regularisation" better here and why it is done.} \lvekhidden{Can you give references here, eg.. to work by other teams such as MWA and LOFAR?}

\subsubsection{Direction-dependent calibration and sky-model subtraction}
\label{sec:dd_cal}

LOFAR has a wide field-of-view (about 10$\degree$ between nulls at 140 MHz,~\citealt{vanHaarlem13}) and the visibilities are susceptible to direction-dependent gain variations mainly due to time varying primary beam and ionospheric effects. Therefore, source subtraction is not a simple deconvolution problem and has to be done with the appropriate gain corrections applied along different source directions. Solving for the gains in each direction would be impractical. The extent of the problem is reduced by (i) clustering the sky-model components~\citep{Kazemi13} in a limited number of directions (here we use 122 directions), (ii) constraining the per-sub-band (195.3 kHz) solutions to be spectrally smooth over the 13\,MHz bandwidth. The number of clusters, which are typically $1-2$ degrees in diameter, is a trade-off between maximizing the S/N inside each cluster and minimizing the cluster size in which all direction-dependent effects (DDE) are assumed to be constant. Constraining the solutions to be spectrally smooth is possible because the earlier direction-independent calibration has taken out most non-smooth instrumental response from the signal chain, and we assume the DDE to be spectrally smooth. We again use \texttt{Sagecal-CO}~\citep{Yatawatta16} with a third-order Bernstein polynomial frequency regularization over the 13 MHz bandwidth to solve for the direction-dependent full Stokes gains, represented by a complex 2$\times$2 Jones matrix~\citep{Hamaker96}. They incorporate all DDE (at this stage mainly the temporally-slow primary beam and ionospheric phase fluctuations). The solution time intervals are chosen between 2.5 and 20 minutes, depending on the apparent total flux in each cluster. This should be adequate for capturing primary beam changes over time, but not for the fast ionospheric phase variations on most baselines~\citep{Vedantham16}. In the future, we plan to investigate the reduction of this solution time interval and to decouple the phase and amplitude solution time~\citep[e.g][]{vanWeeren16}.

\lvekhidden{Define how wide, e.g. between nulls} 
\lvekhidden{Make clear this is only for the DI structure where the gains are applied. DD-structure could still be there is it's not part of the model.}
\lvekhidden{Explain *exactly* what we do with the Jones matrices and whcih is fixed and left free, since this is different from Patil et al. Also what penalty function is used in the optimization? Onlu Stokes I and V, not Q and U?}
\lvekhidden{Reference Harish' and my work on the ionosphere here what these coherences are calculated and defined.}

\lvekhidden{Not clear what you mean here. Do you mean using B-pol usage, or using the same gain for channels insides one sub-band. Try be clear.}

\texttt{Sagecal-CO} uses a consensus optimization with an alternating direction method of multipliers (ADMM) algorithm to efficiently solve for all clusters and all sub-bands simultaneously. The gain solution is constrained to approach a smooth curve by a regularisation prior. As for DI-calibration, here we again use the Bernstein polynomial basis function. \lvekhidden{Make clear here at if one iterates the solution will exactly follow the B-pol, but since we cut short the number of iterations this will not be exactly the case. The regularisation only helps to speeds up the process, but formally only helps the process to converge to an exact B-pol solution.} We use a total of 40 ADMM iterations, which we found to be sufficient to achieve the required convergence. The regularization parameter must be carefully chosen for the fitting process to converge while still enforcing sufficient smoothness. Low or no regularization will effectively over-fit the data, resulting in signal suppression at the smallest baselines where we are most sensitive to the 21-cm signal~\citep{Patil16}. The solution adopted in~\cite{Patil17} is to split the baseline set into non-overlapping calibration and 21-cm signal analysis sub-sets. We chose to exclude the baselines $< 250\,\lambda$ in DD-calibration. This limit is chosen as a compromise: (i) the lower set includes the baselines lengths where we are most sensitive to the 21-cm signal, (ii) it excludes from the calibration the baselines at which the Galactic diffuse emission, not included in our sky-model, starts to be significant, (iii) it still includes enough baselines in the calibration to reach the required S/N. The downside is that the calibration errors now cause excess noise for the baselines not part of the calibration (an effect that was investigated in detail in~\citealt{Patil16}). This additional source of noise can be mitigated by adequately enforcing spectrally smooth solutions, which has the combined benefit of reducing calibration errors, improving the convergence rate, and smoothing the remaining calibration errors along the frequency direction~\citep{Yatawatta15,Barry16}. \cite{Sardarabadi19} have theoretically quantified the level of the expected signal suppression and leakage from direction-dependent calibration. By excluding the $< 250\,\lambda$ baselines during calibration and enforcing spectral smoothness of the gains, they found no signal loss on the baselines of interest and limited amplification for $k_{\parallel}$ modes below $0.15\,\mathrm{h\,cMpc^{-1}}$. Even when considering sky-model incompleteness and that spectral smoothness is only partially achieved, very limited suppression of maximally 5\% is observed. We confirm these results experimentally (Mevius~et~al.~in~prep.) using signals injected in to the data and a setup  identical to our observational and processing setup.


\lvekhidden{Can we motivate this value of the baseline cut? Balance between lost signal to noise and where diffuse emission is low enough.}
\lvek{References here, also to work by others such as MWA, etc.} \fms{Any other than those two that you might think of?}

The regularisation parameters and number of iterations adopted in~\cite{Patil17} were later found to be sub-optimal: the convergence was never reached, resulting in relatively high excess noise. For the analysis presented here, significant focus is placed on improving this aspect. We tested increasing regularization values over a limited set of visibilities (about 1 hour of data) by evaluating the ADMM residuals after each iteration to assess the convergence and gain signal-to-noise ratio. The latter is calculated for every gain-direction (hence cluster of sky-model components) individually and is defined as the ratio of the mean of the gain solution over the standard deviation of the sub-band gain differences. For each individual cluster, we select the regularization value that maximizes the above-mentioned ratio~(Mevius~et~al.~in prep.). Compared to~\cite{Patil17} this ratio is improved by a factor of five. For most clusters, we now reach a S/N ratio $\gtrsim 20$, with clusters inside the first lobe of the primary lobe closer to an S/N ratio of 100 or above~(Mevius~et~al.~in prep.).

\lvekhidden{This regularisation is there only to guide the convergence and hence increased excess noise must have meant that for a fixed number of iterations the solutions are still further away from being smooth. We need to make this clear, since it's fundamentally different from the usual regularisation.}
\lvekhidden{Can we show a plot to illustrate this? Or will this in Maaijke's paper?}
 
Gain-corrected sky-model visibilities are computed after DD-calibration by applying the gain solutions to the predicted sky-model visibilities for each cluster, and subsequently subtracting these from the observed visibilities. Fig.~\ref{fig:clean_img} (bottom panels) shows residual images of the NCP field after DD-calibration. While most of the sources have been correctly subtracted, the brightest sources leave residuals with flux between -50 mJy and +50 mJy.

\subsubsection{Imaging after sky-model subtraction}
\label{sec:imaging}

Residual visibilities obtained after calibration and source subtraction are gridded and imaged independently for each sub-band using \texttt{WSClean}\footnote{https://sourceforge.net/projects/wsclean/}~\citep{Offringa14}, creating an $(l, m, \nu)$ image cube. Recently, several studies analysed the impact of visibility gridding on the 21-cm signal power spectra.~\cite{Offringa19a} assessed the impact of missing data due to RFI flagging and found that the combination of flagging and averaging causes tiny spectral fluctuations, resulting in `flagging excess power' which can be mitigated to a sufficient level by sky-model subtraction before gridding and by using unitary weighted visibilities during gridding\footnote{All visibilities that go into one $uv$-cell are assumed to have the same noise and therefore the same weight.}. The impact of the gridding algorithm itself is also assessed in~\cite{Offringa19b}, and a minimum requirement on various gridding parameters is prescribed. In the present work we follow all these recommendations: (i) our sky-model is subtracted by \texttt{Sagecal} before gridding, (ii) we use unit weighting during gridding, (iii) we use a Kaiser-Bessel anti-aliasing filter with a kernel size of 15 pixels and an oversampling factor of 4095, along with 32 w-layers. These ensure that any systematics due to gridding are confined significantly below the predicted 21-cm signal and thermal noise (see Fig. 8 in~\citealt{Offringa19a} and Fig. 5 in~\citealt{Offringa19b}).

\lvekhidden{Can you provide some details on our gridding settings. $uv$-cell size, etc?}
\lvekhidden{Details on the size of the image, bandwidth of the cubes, resolution of voxels?}
\lvekhidden{Reference Andre's paper here and the figure where this is shown.}
\lvekhidden{Can you indicate the total flux in Jy that is subtracted?}
\lvekhidden{Explain what unit weighting is; is this similar the natural weighting if the errors are the same?}

Stokes I and V images in Jy\,PSF$^{-1}$ and point-spread function (PSF) maps are produced with natural weighting for each sub-band separately.\lvekhidden{Provide more details on the size of the images, pixel resolution, frequency resolution, band-width, etc. Again these numbers need summarising in the table as in Patil et al.} We also create even and odd 10 seconds time-step images to generate gridded time-difference visibilities, which are used to estimate the thermal noise variance in the data. We then combine the different sub-bands to form image cubes with a field of view of 12$\degree\times$12$\degree$ and $0.5\arcmin$ pixel size and these are subsequently trimmed using a Tukey (i.e.\ tapered cosine) spatial filter with a diameter of $4\degree$. This ensures that we reduce our analysis to the most sensitive part of the primary beam, which has a FWHM at 140 MHz of $\approx 4.1\degree$, and avoid the uncertainties of the primary beam at a substantial distance from the beam centre. We choose a Tukey window as a compromise between avoiding sharp edges when trimming the images and maximising the observed volume (i.e.\ maximising the sensitivity).

\lvekhidden{Mention also here comparison to Stokes V?}

\subsection{Conversion to brightness temperature and the combination of power spectra}
\label{sec:gridded_vis_processing}

Here we discuss how visibilities are converted to brightness temperature and how data is averaged both per night of observations and between nights. 

\subsubsection{Conversion to brightness temperature}

The image cube produced by \texttt{WSClean}, $I^{\mathrm{D}}(l, m, \nu)$, has units of Jy/PSF and needs to be converted to units of Kelvin before generating the power spectrum. In order to do that, we recall that the image cube is the spatial Fourier transform of the gridded (and $w$-corrected) visibilities $V_J(u, v, \nu)$, in units of Jansky, with weights $W(u, v, \nu)$ that depend on the chosen weighting scheme~\citep{Thompson01}:
\begin{equation}
\label{eq:dirty_img}
I^{\mathrm{D}}(l, m, \nu) = \sum_{u, v}V_J(u, v, \nu) W(u, v, \nu) e^{+2 \pi i (ul + vm)},
\end{equation}
while the corresponding synthesis beam (or PSF) is given by:
\begin{equation}
\label{eq:dirty_psf}
I^{\mathrm{PSF}}(l, m, \nu) = \sum_{u, v}W(u, v, \nu) e^{+2 \pi i (ul + vm)}.
\end{equation}
Converting the image cube to units of Kelvin consists of dividing out the PSF, i.e. dividing Eq.~\ref{eq:dirty_img} by Eq.~\ref{eq:dirty_psf} in visibility space and converting the measurements to units of Kelvin:
\begin{equation}
T(l, m, \nu) = \frac{10^{-26}c^2}{2k_{\mathrm{B}}\nu^2 \delta_l \delta_m}
\mathcal{F}_{u,v}^{-1} [
\mathcal{F}_
{l,m}[I^{\mathrm{D}}] \oslash \mathcal{F}_{l,m}[I^{\mathrm{PSF}}]],
\end{equation}
with $\mathcal{F}_{l,m}$ denoting the Fourier transform which converts images to visibilities, $\mathcal{F}^{-1}_{u,v}$ its inverse, $k_{\mathrm{B}}$ the Boltzmann constant, $(\delta_l,\,\delta_m)$ the image pixel resolution in radians and $\oslash$ the element-wise division operator.

For each analysed data set, we store the gridded visibilities $V(u, v, \nu)$ in HDF5 format in units of Kelvin, along with the numbers of visibilities that went into each $(u, v, \nu)$ grid point, $N_{\mathrm{vis}}(u, v, \nu)$.

\subsubsection{Outlier flagging}

We use a $k$-sigma clipping method with de-trending, to flag outliers in the gridded visibility cubes. These are likely due to low-level RFI not flagged by \texttt{AOflagger} or due to non-converged gain solutions. Sub-band outliers are flagged based on their Stokes-V and Stokes-I variance, while ($u$, $v$) grid outliers are flagged based on their Stokes-V and sub-band-difference Stokes-I variance. Depending on the dataset, we found that about 20-35\% of the sub-bands and about 5-10\% of $uv$-cells are flagged. At this stage, we are very conservative in our approach to flagging data, favoring less data rather than bad data. These ratios could be reduced in the future by improving low-level RFI flagging before visibilities gridding, and using new algorithms able to filter certain type of RFI instead of flagging them.\lvekhidden{This is quite a lot of data. We need to discuss why it is this high and what is the reason.}

\subsubsection{Noise statistics and weight estimates}
\label{sec:noise_weights}

Several noise metrics are computed to analyse the noise statistics in the data. In general, the noise can be estimated with reasonable accuracy from the Stokes V image cube (circularly polarized sky), the sky being only weakly circularly polarized. Ten second time-difference visibilities, $\delta_t V(u, v, \nu)$, are obtained from taking the difference between the odd and even gridded visibilities sets, yielding a good estimate of the thermal noise (at this time resolution, the foregrounds and ionospheric errors cancel out almost perfectly). We can compare it to the per-station system equivalent flux density (SEFD), given that the gridded visibility thermal-noise rms $\sigma(u, v, \nu)$ follows, by definition~\citep{Thompson01},
\begin{equation}
\label{eq:sefd}
\sigma(u, v, \nu) = \frac{1}{N_{\mathrm{vis}}(u, v, \nu)}\frac{\mathrm{SEFD}}{
\sqrt{2 \Delta \nu \Delta t}},
\end{equation}
with $\Delta \nu$ and $\Delta t$ the frequency channel and integration time, respectively. Using Eq.~\ref{eq:sefd}, we estimate the SEFD of the 12 nights analysed to be $\approx 4150$ Jy (almost constant over the the 13 MHz bandwidth) \lvekhidden{At what frequency?} with a standard deviation of $\approx 160$ Jy (fifth column of Table~\ref{tab:all_nights}). This is similar to the empirical values estimated in~\cite{vanHaarlem13} for the LOFAR-HBA core stations, after correction for the primary beam sensitivity in the direction of the NCP~\citep{Patil17}. The small night-to-night variation could be attributed to a combination of different observing LST time (the sky noise being one component of the thermal noise, along with the system noise) and/or missing tiles for some of the stations during some nights.\lvek{What about different levels of flagging and that some nights are longer than others, 12-16hr?}\fms{this is taken into account in $N_{\mathrm{vis}}$} We also note that our absolute calibration is accurate at the 5\% level.

\begin{figure}
    \includegraphics{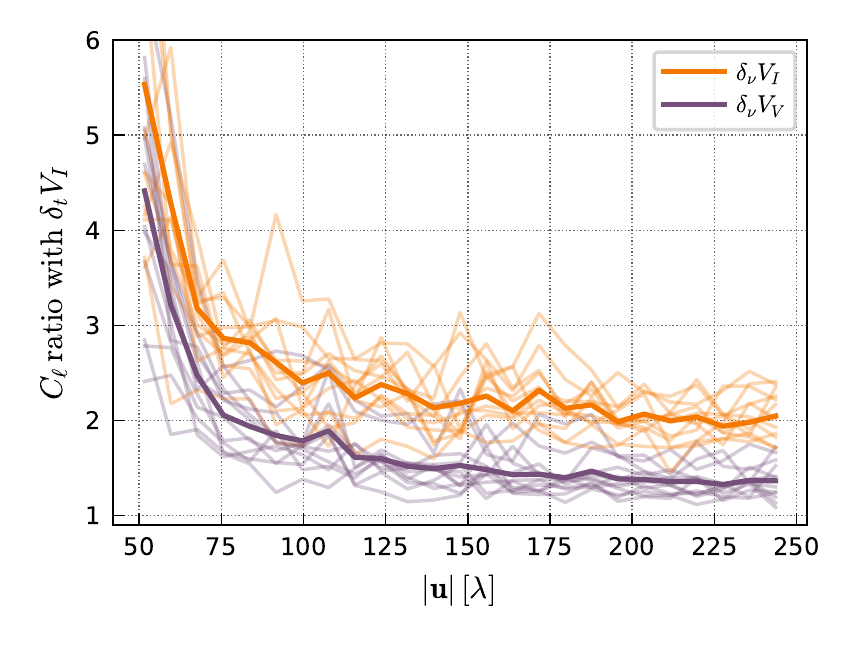}
    \caption{\label{fig:ps_ratio_dt_df} Ratio between sub-band difference and time difference angular power spectra for Stokes I (orange lines) and Stokes V (magenta lines). All nights are shown, and the average over all nights is indicated by thicker line.}
\end{figure}

Another noise estimate can be derived from the visibility difference between sub-bands, $\delta_ {\nu} V(u, v, \nu)$, which should better reflect the spectrally-uncorrelated noise in the data. Compared to the time difference noise spectrum (in baseline-frequency space), we find that the sub-band difference noise variance is on average higher by a factor $\approx 1.35$ for Stokes V and $\approx 2$ for Stokes I (sixth and seventh columns of Table~\ref{tab:all_nights}, respectively) with a small night-to-night variation. We also find that this additional spectrally-uncorrelated noise term is dependent on the baseline length, with the ratio of the sub-band difference over time difference noise spectrum gradually increasing as a function of decreasing baseline length. A similar trend is observed for both Stokes I and V (see Fig.~\ref{fig:ps_ratio_dt_df}).

\begin{figure}
    \includegraphics{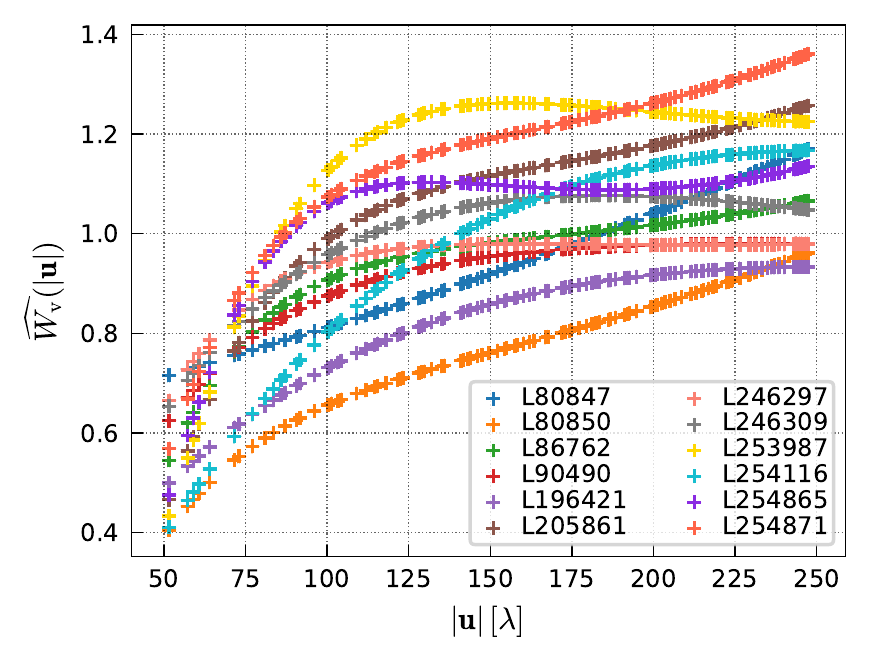}
    \caption{\label{fig:weights_all_nights} Weights scaling factor $\widehat{W_\mathrm{v}}$ as a function of baseline length, for all nights (one color per night).}
\end{figure}

While the origin of this increased noise is still being investigated, and will be discussed in more detail in Section~\ref{sec:results_individual}, it needs to be taken into account when weighting the data. Inverse variance weighting is used to obtain an optimal average over the data sets from different nights and for power spectrum estimation. Theoretically, if all visibilities had the same noise statistics, the optimal thermal-noise weights would be given by the effective number of visibilities that went inside each $(u, v)$ grid point, $N_ {\mathrm{vis}}(u, v, \nu)$. Here, we additionally account for the night-to-night and baseline variation of the noise using Stokes-V sub-band difference noise estimates by computing:
\begin{equation}
W_\mathrm{v}(u, v) = \frac{1}{\mathrm{MAD}_{\nu}(\delta_{\nu} V_V(u, v, \nu)
\sqrt{N_{\mathrm{vis}}(u, v, \nu)})^2}
\end{equation}
with MAD denoting the median absolute deviation estimator. This effectively computes weights based on per-visibility Stokes V variance which we then combined with the weights related to the $(u, v)$ density of the gridded visibilities. The per-visibility noise variance is theoretically invariant and any night-to-night or baseline-dependent variation will be reflected in $W_\mathrm{v}$. Because we are mainly interested in accounting for the baseline variation of the noise, we additionally perform a third-order polynomial fit of $W_\mathrm{v}(|\mathbf{u}|)$ to form $\widehat{W_\mathrm{v}}(|\mathbf{u}|)$, and a normalization such that $\left<\widehat{W_\mathrm{v}}(| \mathbf{u}|)\right> = 1$ averaged over all nights and all baselines. This makes this estimator even more robust against outliers and biases due to small number statistics. The final weights per night are then given by:
\begin{equation}
\label{eq:weights}
W(u, v, \nu) = N_{\mathrm{vis}}(u, v, \nu) \widehat{W_\mathrm{v}}(|
\mathbf{u}|).
\end{equation}
The scaling factor $\widehat{W_\mathrm{v}}(|\mathbf{u}|)$ for all nights is plotted in Fig.~\ref{fig:weights_all_nights}.
\lvekhidden{A figure might help to illustrated this.}

\subsubsection{Averaging multiple nights}
\label{sec:nights_averaging}

It is necessary to combine several nights of observation to reduce the thermal noise level. It is expected that a total of about 1000 hours of LOFAR-HBA observation on one deep field will be required for a statistical detection of the 21-cm signal from the EoR. In the present work, 12 nights are analysed, of which the best 10 nights are combined, totalling 141 hours of observations. The different nights are combined in visibilities with the weights obtained from Eq.~\ref{eq:weights}:
\begin{equation}
 V_{\mathrm{c}n}(u, v, \nu) = \frac{\sum^{n}_{i=1} V_i(u, v, \nu) W_i(u, v,
 \nu)}{\sum^{n}_{i=1} W_i(u, v, \nu)}
\end{equation}
where $V_i$ is the visibility cube of the $i$-th night, and $V_{\mathrm{c}n}$ is the visibility cube of $n$ nights combined.

\subsection{Residual foreground removal}
\label{sec:fg_removal}

After direction-dependent calibration and subtraction of the gain-corrected sky model, the residual Stokes-I visibilities are composed of extragalactic emission below the confusion limit (and thus not removable by source subtraction) and partially-polarized diffuse Galactic emission which is still approximately three orders of magnitude brighter than the 21-cm signal. The emission mechanism of these foreground sources (predominantly synchrotron and free-free emission) are well-known to vary smoothly in frequency, and this characteristic can differentiate them from the rapidly fluctuating 21-cm signal~\citep{Shaver99,Jelic08}. However, the interaction of the spectrally smooth foregrounds with the Earth's ionosphere, the inherent chromatic nature of our observing instrument (in both the PSF and the primary beam), and chromatic calibration errors create additional `mode-mixing' foreground contaminants which introduce spectral structure to the otherwise smooth foregrounds~\citep{Datta10,Morales12,Trott12,Vedantham12}. 

In the two-dimensional angular ($k_{\perp}$) versus line-of-sight ($k_{\parallel}$) power spectra, the foregrounds and mode-mixing contaminants are primarily localized inside a wedge-like region\footnote{This peculiar shape is explained by the fact that longer baselines (higher $k_{\perp}$) change length more rapidly as a function of frequency than smaller baselines, causing increasingly faster spectral fluctuations, and thus producing power into proportionally higher $k_{\parallel}$ modes.}. This makes them separable from the 21-cm signal by either avoiding the predominantly foreground-contaminated region and only probe a $k$-space region where the 21-cm signal dominates (foreground avoidance strategy; e.g.~\citealt{Liu14a,Trott16b}), or by exploiting their different spectral (and spatial) correlation signature to separate them (foreground removal strategy; e.g.~\citealt{Chapman12,Chapman13,Patil17,Mertens18}). 

We adopt a foreground removal strategy which, if done correctly, has the advantage of considerably increasing our sensitivity to larger co-moving scales (smaller $k$-modes)~\citep{Pober14}. To that aim, we developed a novel foregrounds removal technique based on Gaussian Process Regression (GPR)~\citep{Mertens18}. In this framework, the different components of the observations, including the astrophysical foregrounds, mode-mixing contaminants, and the 21-cm signal, are modelled as a Gaussian Process (GP). A GP is the joint distribution of a collection of normally distributed random variables~\citep{Rasmussen05}. The sum of the covariances of these distributions, which define the covariance between pairs of observations (e.g.\ at different frequencies), is specified by  parameterizable covariance functions. The covariance function determines the structure that the GP will be able to model. In GPR, we use the GP as parameterized priors, and the Bayesian likelihood of the model is estimated by conditioning this prior to the observations. Standard optimization or MCMC methods can be used to determine the optimal hyper-parameters of the covariance functions. The GPR method is closely related to Wiener filtering~\citep{Zaroubi95,Sarkka13}. Compared to the Generalized Morphological Component Analysis~\citep[GMCA,][]{Bobin07,Chapman13} used in~\cite{Patil17}, GPR is more suited to treat the problem of foregrounds in high redshift 21-cm experiments~\citep{Mertens18} and reduces the risk of signal suppression by explicitly incorporating a 21-cm signal covariance prior in its GP covariance model. 


\subsubsection{Gaussian Process Regression}

\lvekhidden{I made the text below a subsubsection, but I think the extensive math might detract from the flow of the paper and would suggest we move this to an appendix and only leave the above description here. The method has been described several times already. The only relevant parts that could stay in the main body of the paper are the actual choices for the covariance functions, their parameters priors. These need to go again in to a summarising table as in Patil et al.}

Formally, we model our data $\mathbf{d}$ observed at frequencies $\mathbf{\nu}$ by a foreground $\mathbf{f}_{\mathrm{fg}}$, a 21-cm signal $\mathbf{f}_{\mathrm{21}} $and noise $\mathbf{n}$ components:
\begin{equation}
\mathbf{d} = \mathbf{f}_{\mathrm{fg}} + \mathbf{f}_{\mathrm{21}} +
\mathbf{n}.
\end{equation}
\lvekhidden{I still think the notation is non-standard according to MNRAS. Using vector and matrices. Also matrices should not be italicized. Please check the MNRAS style to see how equations are written for all the text below. For example vectors should not be written as functions above, which I changed. I tried to adapt everything below, but double check.}

The foreground signal can be statistically separated from the 21-cm signal by exploiting their different spectral behavior. The covariance of our GP model (in GPR the covariance matrix entries are defined by a paramaterised function and the distance between entries in the data-vector, e.g. the difference in frequency) can then be composed of a foreground covariance $\rm K_{\mathrm{fg}}$ and a 21-cm signal covariance $\rm K_{\mathrm{21}}$,
\begin{equation}
\rm K = K_{\mathrm{fg}} + K_{\mathrm{21}}.
\end{equation}
The foreground covariance itself is decomposed into two parts, accounting for the large frequency coherence scale of the intrinsic extragalactic and Galactic foreground emission and the smaller frequency coherence scale (in the range of $1 - 5$ MHz) of the mode-mixing component\footnote{Formally the chromatic nature of the instrument implies that mode-mixing has a multiplicative effect, but this can be approximated, to first order, as an additive effect, justifying the use of separable additive covariance for large and small frequency coherence scale foregrounds.}.

\lvekhidden{Notation. These are matrices. Switch early to that notation to avoid confusion. Simply state that the covariance matrix entries are defined by a paramaterised function and the distance between entries in the data-vector, e.g. the difference in frequency.}

We use an exponential covariance function for the 21-cm signal, as we found that it was able to match well the frequency covariance from a simulated 21-cm signal~\citep{Mertens18}. Eventually, the choice of the covariance functions is data-driven, in a Bayesian sense, selecting the one that maximizes the evidence. We will see in Section~\ref{sec:results_individual} that the simple foregrounds + 21-cm dichotomy will need to be adapted, introducing an additional component, to match the data better.

The joint probability density distribution of the observations $\mathbf{d}$ and the function values $ \mathbf{f}_{\mathrm{fg}}$ of the foreground model at the same frequencies $\nu$ are then given by,
\begin{equation}
\left[ \begin{array}{c} \mathbf{d} \\ \mathbf{f_{\mathrm{fg}}} \end{array}\right] \sim  
\mathcal{N}\left( \left[\begin{array}{c} 0 \\ 0 \end{array}\right], \left[ 
\begin{array}{cc} \rm K_{\mathrm{fg}} + K_{\mathrm{21}} + \rm K_{\mathrm{n}} & \rm K_{\mathrm{fg}} \\ \rm K_{\mathrm{fg}} & \rm K_{\mathrm{fg}} \end{array} \right] \right)
\end{equation}
using the shorthand $\rm K \equiv K(\nu, \nu)$, and where $\rm K_{\mathrm{n}} = \mathrm{diag}(\sigma_{ \mathrm{n}}^2(\nu))$ is the noise covariance. The foreground model is then a Gaussian Process, conditional on the data:
\begin{equation}
\label{eq:gpr_fg_fit_distribution}
\mathbf{f}_{\mathrm{fg}} \sim \mathcal{N}\left({\cal E}(\mathbf{f}_{\mathrm{fg}}), \mathrm{cov}(\mathbf{f}_{\mathrm{fg}})\right)
\end{equation}
with expectation value and covariance defined by:
\begin{align}
\label{eq:gpr_predictive_mean_eor} {\cal E}(\mathbf{f}_{\mathrm{fg}}) &= \rm K_{\mathrm{fg}}\left[K_{\mathrm{fg}} + K_{\mathrm{21}} + K_{\mathrm{n}}\right]^{-1} 
\mathbf{d}\\
\label{eq:gpr_predictive_cov_eor}
\mathrm{cov}(\mathbf{f}_{\mathrm{fg}}) &\rm = K_{\mathrm{fg}} - K_{
\mathrm{fg}}\left[K_{\mathrm{fg}} + K_{\mathrm{21}} + K_{\mathrm{n}}\right]^{-1}K_{\mathrm{fg}}.
\end{align}
The residual is obtained by subtracting ${\cal E}(\mathbf{f}_{\mathrm{fg}})$ from the observed data:
\begin{equation}
\mathbf{r} =\mathbf{d} - {\cal E}(\mathbf{f}_{\mathrm{fg}}).
\end{equation}
\begin{figure*}
    \includegraphics{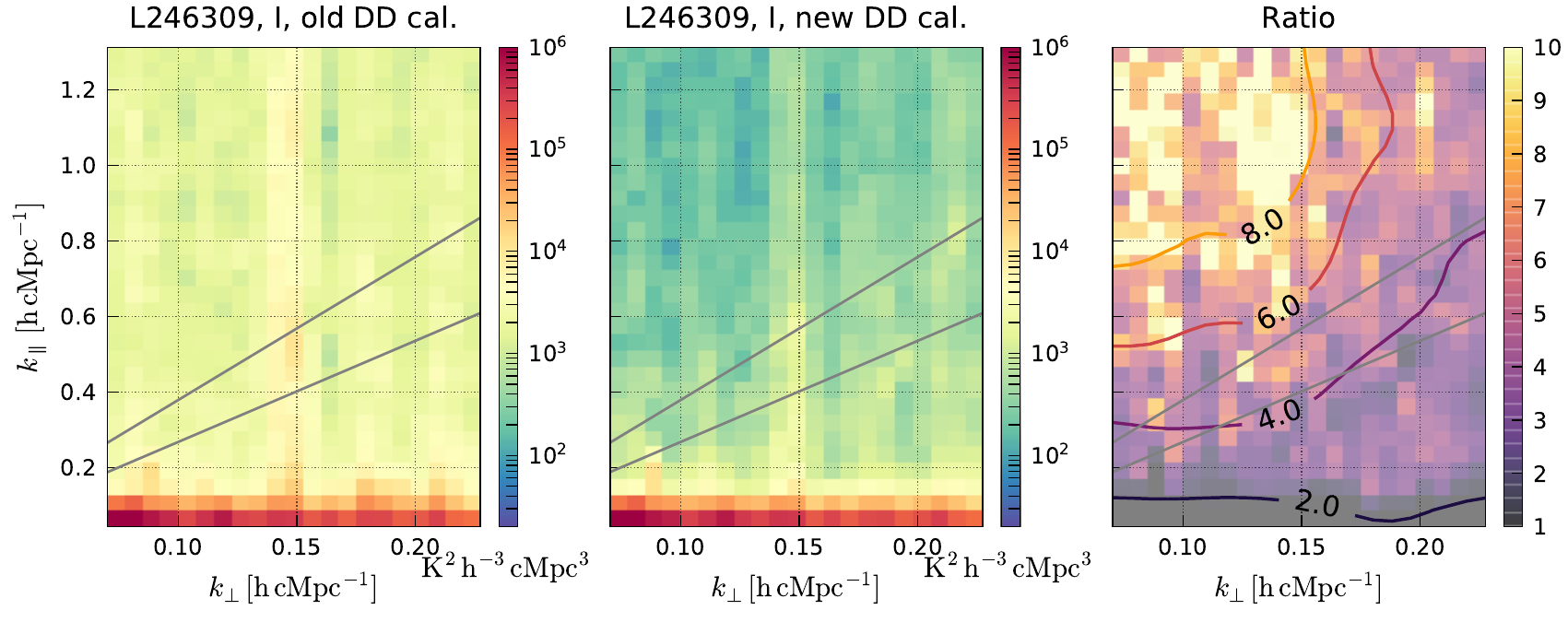}
    \caption{\label{fig:ps2d_L246309_I_low_reg_vs_high_reg} Improvement due to the new calibration for a single night of observation. We compare the new DD-calibration procedure (middle panel) against the one adopted in~\protect\cite{Patil17} (left panel). The ratio of the two (right panel) shows a substantial reduction of the excess noise related to the $250\lambda$ baseline cut over-fitting effect (by a factor $> 5$ for $k_{\parallel} > 0.8 \mathrm{h\,cMpc^{-1}}$), with no impact on the residual foregrounds (ratio $\sim 1$ at low $k_{\parallel}$). The plain gray lines indicate, from bottom to top, $50\degree$ and instrumental horizon delay lines (delimiting the foreground wedge).}
\end{figure*}

\begin{figure}
    \includegraphics{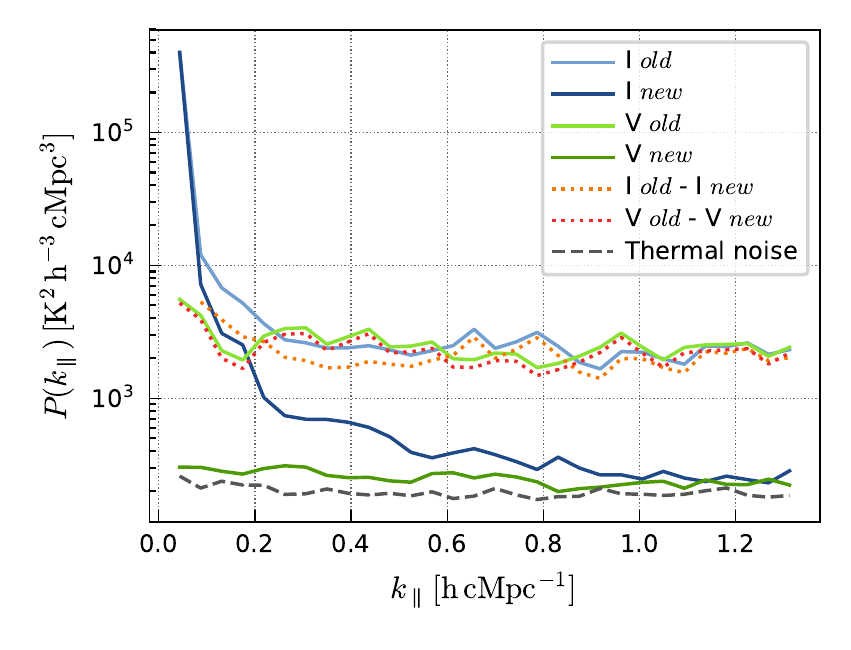}
    \vspace{-0.5cm}
    \caption{\label{fig:ps2d_kpar_L246309_I_V_low_reg_vs_high_reg} Improvement due to
    the new calibration for a single night of observation. Here we compare Stokes I (blue
    lines) and Stokes V (green lines) cylindrically averaged power spectra
    (averaged over all baselines) processed with the new DD-calibration
    procedure (\textit{new}) against the one used in
    ~\protect\cite{Patil17} (\textit{old}). The excess noise (difference between
    \textit{old} and \textit{new}) is reduced similarly in Stokes I (orange
    line) and Stokes V (red line). The thermal noise power is indicated by the
    dashed gray line.}
\end{figure}

\subsubsection{Bias corrections}

Inferring the variance of a distribution in general leads to a bias when its expectation value is also inferred at the same time. \fms{would you suggest any good reference here ?} To correct for this bias, we derive an unbiased version of the residual covariance (or power spectra). The residual covariance is formally given by:
\begin{equation}
\langle \mathbf{r}\, \mathbf{r}^{\rm H} \rangle = \langle (\mathbf{d} - {\cal E}(\mathbf{f}_{\mathrm{fg}}))(\mathbf{d} - {\cal E}(\mathbf{f}_{\mathrm{fg}}))^{\rm H} \rangle
\end{equation}
which, after replacing ${\cal E}(\mathbf{f}_{\mathrm{fg}})$ by Eq.~\ref{eq:gpr_predictive_mean_eor}, and introducing the residual covariance $\rm K_r = K_{21} + K_{\mathrm{n}}$, evaluates to:
\begin{equation}
\begin{split}
\langle \mathbf{r}\, \mathbf{r}^{\rm H} \rangle &\rm = (I - K_{\mathrm{fg}}[K_{\mathrm{fg}} + K_{\mathrm{r}}]^
{-1}) \langle \mathbf{d}\, \mathbf{d}^{\rm H}\rangle \\
&\rm \qquad (I - [K_{\mathrm{fg}} + K_{\mathrm{r}}]^{-1}K_
{\mathrm{fg}}).
\end{split}
\end{equation}
Assuming the GP covariance model is adequate (which translates to ${<\mathbf{d} \, \mathbf{d}^{\rm H}> = \rm K_{\mathrm{fg}} + K_{\mathrm{r}}}$), we have:
\begin{align}
\begin{split}
\langle \mathbf{r}\, \mathbf{r}^{\rm H}\rangle &= \rm (I - K_{\mathrm{fg}}[K_{\mathrm{fg}} + K_{\mathrm{r}}]^{-1}) (K_{
\mathrm{fg}} + K_{\mathrm{r}}) \\
& \rm \qquad(I - [K_{\mathrm{fg}} + K_{\mathrm{r}}]^{-1}K_{\mathrm{fg}}) \nonumber
\end{split}
 \\
&\rm =K_{\mathrm{r}} - K_{\mathrm{r}}[K_{\mathrm{fg}} + K_{\mathrm{r}}]^{-1}K_{
\mathrm{fg}} \nonumber \\
\begin{split}
&\rm =K_{\mathrm{r}} - (K_{\mathrm{fg}} + K_{\mathrm{r}})[K_{\mathrm{fg}} + K_{
\mathrm{r}}]^{-1}K_{\mathrm{fg}} \\
&\rm \qquad + K_{\mathrm{fg}}[K_{\mathrm{fg}} + K_{\mathrm{r}}]^{-1}K_{\mathrm{fg}} 
\nonumber 
\end{split}
\\
&\rm =K_{\mathrm{r}} - K_{\mathrm{fg}} + K_{\mathrm{fg}}[K_{\mathrm{fg}} + K_{
\mathrm{r}}]^{-1}K_{\mathrm{fg}} \nonumber \\
&\rm =K_{\mathrm{r}} - \mathrm{cov}(\mathbf{f}_{\mathrm{fg}}).
\end{align}
We see that, in order to obtain the expected covariance of the residual, $\rm K_{\mathrm{r}}$, we need to un-bias the estimator using $\mathrm{cov}(\mathbf{f}_{\mathrm{fg}})$. An unbiased estimator of the covariance of the residual is then given by:
\begin{equation}
\langle \mathbf{r}\, \mathbf{r}^{\rm H} \rangle_{\mathrm{unbiased}} = \langle (\mathbf{d} - {\cal E}(\mathbf{f}_{\mathrm{fg}}))(\mathbf{d} - {\cal E} (\mathbf{f}_{\mathrm{fg}}))^{\rm H}\rangle + \mathrm{cov}(\mathbf{f}_{\mathrm{fg}}).
\end{equation}
Intuitively, this can be understood by considering that ${\rm E}(\mathbf{f}_{\mathrm{fg}})$ is just one possible realization of the foreground fit (the maximum a-posterior, i.e.\ MAP, solution), and any function derived from the distribution defined in Eq.\,\ref{eq:gpr_fg_fit_distribution} is a valid foreground fit to the data. Similar derivations can be obtained for the power spectra. The above bias correction has been tested numerically.

\lvekhidden{Explain what $\bf f$ is. Its not been defined yet. Is it f of FG?}

\subsection{Power spectra estimation}

Given the observed brightness temperature of the
21-cm signal $T(\mathbf{r})$ as a function of spatial coordinate $\mathbf{r}$, the
power spectrum $P(\mathbf{k})$ as a function of wavenumber $\mathbf{k}$ is defined
as:
\begin{equation}
P(\mathbf{k}) = \mathbb{V}_c |\tilde{T}(\mathbf{k})|^2,
\end{equation}
with $\tilde{T}(\mathbf{k})$ the discrete Fourier transform of the temperature field
defined as:
\begin{equation}
\tilde{T}(\mathbf{k}) = \frac{1}{N_l N_m N_\nu} \sum_{\mathbf{r}} T(\mathbf{r})
e^{-2 i \pi \mathbf{k} \mathbf{r}},
\end{equation}
and $\mathbb{V}_c$ is the observed comoving cosmological volume, delimited by
the primary beam of the instrument $A_\mathrm{pb}(l, m)$, the spatial tapering
function $A_w(l, m)$ and frequency tapering function $B_w (\nu)$ applied to the
image cube before the Fourier transform:
\begin{align}
&\mathbb{V}_c = \frac{(N_l N_m N_{\nu} dl dm d\nu) \mathrm{D}_M(z)^2 \mathrm{\Delta
D}} {A_{\mathrm{eff}} B_{\mathrm{eff}}} \\
&A_{\mathrm{eff}} = \langle A_\mathrm{pb}(l, m)^2A_w(l, m)^2\rangle \\
&B_{\mathrm{eff}} = \langle B_\mathrm{w}(\nu)^2 \rangle.
\end{align}
Here $\mathrm{D}_M(z)$ and $\mathrm{\Delta D}$ are conversion factors from angle
and frequency, respectively, to comoving distance. We also define the wavenumber
$\mathbf{k} = (k_l, k_m, k_{\parallel})$ as~\citep{Morales04,McQuinn06} \lvekhidden{Ref to McQuinn et al. 2006 and to Morales et al. 2004}:
\begin{equation}
k_{l} = \frac{2 \pi u}{D_M(z)},\,k_{m} = \frac{2 \pi v}{D_M(z)},\,k_{\parallel}
= \frac{2 \pi H_0 \nu_{21} E(z)}{c(1+z)^2} \eta,
\end{equation}
where $H_0$ is the Hubble constant, $\nu_{21}$ is the frequency of the hyperfine transition, and $E(z)$ is the dimensionless Hubble parameter~\citep{Hogg10}. With the assumption of an isotropic signal, we can average $P(\mathbf{k})$ in $k$-bins creating the spherically averaged dimensionless power spectrum defined as:
\begin{equation}
\Delta^2({k}) = \frac{k^3}{2 \pi^2} \left< P(\mathbf{k}) \right>_k.
\end{equation}
For diagnostic purposes, we also generate the variance of the image cube as a function of frequency, cylindrically averaged power spectra, and angular power spectra ($C_{\ell}$) which characterize the transverse scale fluctuation average over all frequencies. We define the cylindrically averaged power spectrum, as a function of angular ($k_{\perp}$) versus line-of-sight ($k_{\parallel}$) scales as:
\begin{equation}
P(k_{\perp}, k_{\parallel}) = \left< P(\mathbf{k}) \right>_{k_{\perp}, k_{\parallel}}.
\end{equation}
The angular, spherical and cylindrical power spectra are all optimally weighted using the weights derived in Section~\ref{sec:noise_weights}. The $k_{\parallel} = 0$ modes are discarded from the spherical and cylindrical power spectra calculations as they are considered unreliable for 21-cm signal detection (for these modes, the foregrounds and 21-cm signal are statistically difficult to distinguish).

The uncertainties on the power spectra reported here are sample variance taking into account the number of individual $uv$-cells averaged, and the effective observed field-of-view given by the primary beam $A_\mathrm{pb}(l, m)$ and spatial tapering function $A_w(l, m)$. They assume that all averaged $uv$-cells are independent measurements\footnote{The primary beam and spatial tapering function introduce correlation, but those can be ignore at the scales we measure our power spectra: the width of the primary beam and tapering window is 4 times larger than the scale probed by our smallest baseline of $50\,\lambda$.}. All residual and noise power spectra are computed without a frequency-tapering function to benefit from the full bandwidth sensitivity. In the case of GPR residuals, we have another source of uncertainty which comes from the uncertainty on the GP model hyper-parameters. These can be propagated using an MCMC method (see Appendix~\ref{sec:mcmc}). This calculation shows it to be negligible compared to the sample variance and it can be ignored in our calculations (see also~\citealt{Mertens18}).

\lvekhidden{Is this correlation important or are the chosen $uv$-cell large enough that covariance can be ignored. Please comment on this.} 

Foreground emission is usually confined to a wedge-like structure in $k$ space~\citep{Datta10,Morales12}. This wedge line is defined by:
\begin{equation}
\label{eq:wedge}
k_{\parallel}(\theta; k_{\perp}) = \frac{H_{0}D_{M}(z)E(z)}{c(1+z)} \mathrm{sin}(\theta) k_{\perp},
\end{equation}
where $\theta$ is the angular distance from the phase center of the foreground source. The instrumental horizon delay line is given setting $\theta = 90 \degree$ and delimits the `foreground wedge' ($k_{\parallel}$ modes below this line) and `EoR window' ($k_{\parallel}$ modes above this line) regions.

\begin{figure*}
    \includegraphics{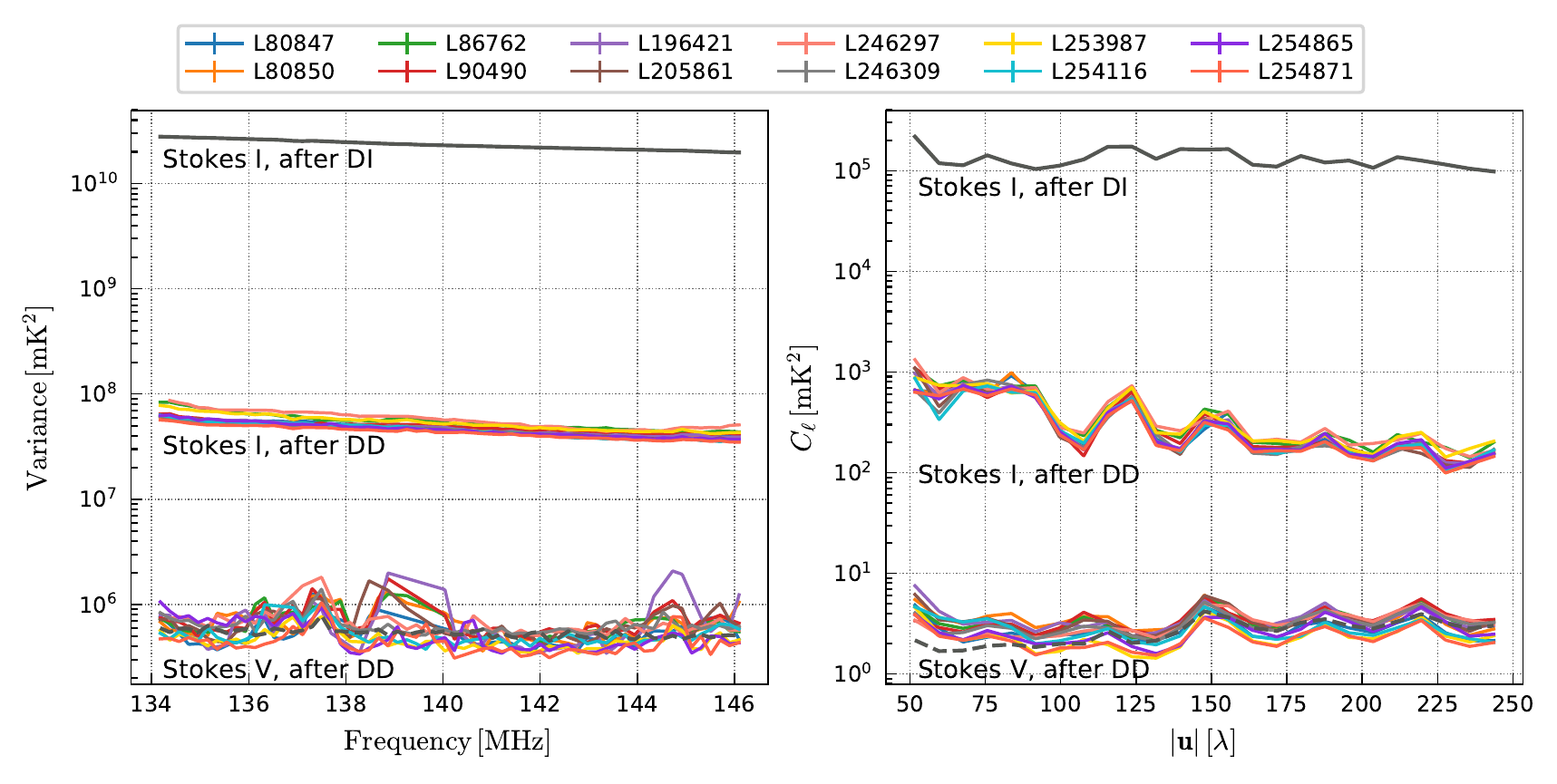}
    \caption{\label{fig:var_cl_all_nights}Variance (left panel) and angular
    power spectra (right panel) for all nights at different processing stages.
    Different nights are indicated by a different color. The top lines show the Stokes
    I power after DI-calibration. The middle lines show the Stokes I power after
    DD-calibration and sky model subtraction (but before GPR). The lines at the bottom show the
    Stokes V sub-band difference power. The black dashed line represent the
    thermal noise power for an average observing duration time (14.4 h) and an
    average SEFD (4150 Jy).}
\end{figure*}

\section{Results from night to night}

\label{sec:results_individual}

In this section we discuss the results of processing the data from each night individually. We start by assessing the improvement made to the data processing compared to~\cite{Patil17}. The residual foregrounds (after DD-calibration) and noise in the data are analysed and we examine the residual image cubes after GPR foreground removal, and its night-to-night correlation.

\subsection{Power spectra before foreground removal}

All nights are calibrated and imaged following the procedure described in
Section~\ref{sec:formalism}. 

\subsubsection{Calibration improvements}

To demonstrate the improvement in the  calibration, we process one night of observation (L246309) with the DD-calibration regularization parameters used in~\cite{Patil17}. Mevius~et~al.~(in~prep) show that the latter approach leads to substantial excess noise (beyond thermal noise), in particular if the constraints on spectral smoothness are not correctly enforced. This leads to excess noise on baselines $< 250\,\lambda$ because of over-fitting (see also~\citealt{Sardarabadi19}). Cylindrically-averaged power spectra of Stokes I and Stokes V for the two calibration procedures (\textit{old} vs \textit{new}) are shown in Figures~\ref{fig:ps2d_L246309_I_low_reg_vs_high_reg} and~\ref{fig:ps2d_kpar_L246309_I_V_low_reg_vs_high_reg}, indicating a significant decrease of the excess noise, while leaving the residual foregrounds largely unaffected. Taking the difference between the \textit{old} and \textit{new} procedures shows that the excess noise is reduced in both Stokes I and Stokes V in a similar manner (see Figure~\ref{fig:ps2d_kpar_L246309_I_V_low_reg_vs_high_reg}). This excess noise is mostly spectrally uncorrelated and close to constant as a function of $k_{\parallel}$, with the small increase of power at $k_{\parallel} < 0.2 \mathrm{h\,cMpc^{-1}}$ related to the basis function adopted as frequency gain constraint. This is in good agreement with the theoretical predictions from~\cite{Sardarabadi19}. With the new procedure, the Stokes V power is now also closer to the thermal noise power.

\lvekhidden{The causal connection to overfitting and uncorrelated noise is not clear. Can you reference some works here? On the "brick"?}

\subsubsection{Residual foregrounds}

Figure~\ref{fig:var_cl_all_nights} shows the total intensity variance and angular power spectra at different steps of processing. The foreground power is reduced by a factor of $\sim 500$ after DD-calibration. The residual power is consistent between nights, with a night-to-night relative variation of $\approx12 \%$. The Stokes-I angular power spectra are relatively flat before sky-model subtraction, while afterwards, the power toward the larger scales (smaller baselines) increases, consistent with a power-law with a spatial slope $\beta_{\ell} \approx-1.18$. On large scales, the observed residual power, $C_{\ell} (|\mathbf{u}| = 50 \lambda) \sim 10^3\,\mathrm{mK^2}$, is comparable with the power attributed to the Galactic foregrounds in the NCP field observation from~\cite{Bernardi10} using the Westerbork telescope. However, the spatial slope does not match the expectation from Galactic diffuse emission, in the range $[-2, -3]$~\citep{Bernardi10}. This suggests that the residual power observed here is a combination of Galactic emission, residual confusion-limited extragalactic sources, and calibration errors from the DD-calibration stage. The latter may be substantial (see e.g~\citealt{Sardarabadi19}), but because they are now mostly frequency coherent (resulting from the high regularization used in the consensus optimization), they are separable from the 21-cm signal and can be removed using the GPR method.

\subsubsection{Noise statistics}
\label{sec:noise_stats}

Following the procedure detailed in Section~\ref{sec:noise_weights}, Stokes-V and Stokes-I sub-band difference power spectra ($\delta_{\nu} I$ and $\delta_{\nu} V$, respectively) are generated as a proxy for spectrally-uncorrelated noise, and time-difference power spectra from even/odd sets are generated as a proxy for the thermal noise power spectra ($\delta_{t} V$). \lvekhidden{You mix power and noise. Be clear on this. If you say noise, do you mean the rms or the power. Maybe say "noise power" instead to make it clear.}
Taking the power ratio of $\delta_{\nu} V$ over $\delta_{t} V$, exhibits a  non-negligible excess power well above the thermal noise level ($\approx\,35\%$, see Table~\ref{tab:all_nights}). This additional spectrally-uncorrelated noise is baseline dependent, with a flat ratio of $\approx\,1.25$ for baselines of length $> 125\,\lambda$, and then gradually increasing to smaller baselines (see Figures~\ref{fig:var_cl_all_nights} and~\ref{fig:ps_ratio_dt_df}). The ratio also varies considerably from night to night.
Examining the power ratio of $\delta_{\nu} V$ over $\delta_{\nu} I$, shows a higher sub-band difference noise level (by a factor $\approx 50\%$) in Stokes I. This ratio has a weak dependence on the baseline length (with a Pearson correlation coefficient between ratio and baselines $r =0.23$ and a corresponding p-value $<10^{-5}$).

This source of noise is still being investigated. One hypothesis is mutual-coupling between spatially close stations~\citep[e.g.][]{Fagnoni19}. This would explain the rise of power with decreasing baseline length. It might also be a source of broadband and faint RFI at the central LOFAR "superterp" region. It is also interesting to note that the Galactic diffuse emission is prominent at baselines $< 125\,\lambda$. Each of these effects will be further analysed in future publications.

\lvekhidden{Discuss origin of the spectrally uncorrelated noise.}

\subsection{Residual foreground removal}
\label{sec:gpr_fg_removal}

The residual foreground emission after DD calibration is removed using GPR modelling which is applied to the same gridded visibilities ($4 \degree \times4 \degree$ field of view) as used for the power spectrum analysis.

\subsubsection{Covariance model}
\label{sec:cov_model}

In Section~\ref{sec:fg_removal} it was shown that we can recover unbiased power spectra of the signal as long as the covariance model matches the data. The GP model therefore needs to be as comprehensive as possible, incorporating covariance functions for all components of the data, including the 21-cm signal and known systematics. The selection of the covariance functions is driven by the data in a Bayesian framework, by selecting the model that maximizes the evidence. Because these covariance functions are parametrized, they too are optimized.

\setlist[description]{font=\normalfont\itshape}

\renewcommand{\arraystretch}{1.2}
\begin{table}
\centering
\caption{Different GP models assessed against the fiducial GP model, being a Matern kernel with $\eta_\mathrm{mix} = 3 / 2$ (see Section~\ref{sec:cov_model}). \lvekhidden{State that model here or refer to it in the text.} Negative values of the difference in log-evidence ($\cal Z$) indicate a less probable model. A difference of $|\Delta {\cal Z}|>20$ is typically regarded as a very strong difference in evidence.}
\begin{threeparttable}
\label{tab:all_gp_test}
\begin{tabular}{lr}
\toprule
Model change & $\Delta {\cal Z}$ \\
\midrule
$\eta_{\mathrm{ex}} = 5 / 2$, $\eta_{\mathrm{mix}} = 3 / 2$ (fiducial) & 0 \\
$\eta_{\mathrm{mix}} = 5 / 2$ & -39 \\
$\eta_{\mathrm{mix}} = +\infty$ & -147 \\
$\kappa_{\mathrm{mix}} \equiv \kappa_{\mathrm{RatQuad}}$ & -7 \\
$\alpha_n = 1$ (fixed) & -110 \\
$\sigma^2_{\mathrm{ex}} = 0$ (fixed) & -149 \\
$\eta_{\mathrm{ex}} = 3 / 2$ & -17 \\
\midrule
\end{tabular}
\end{threeparttable}
\end{table}

\renewcommand{\arraystretch}{1}

\begin{description}
    \setlength\itemsep{1em}
    \item[(1) The foregrounds ---] 
At this stage, the foreground residuals are mainly composed of intrinsic sky emission from confusion-limited extragalactic sources and from our own Galaxy, and of mode-mixing contaminants related to e.g.\ the instrument chromaticity and calibration errors that can originate from all sources in the sky leaking into the $4 \degree \times4 \degree$ image cubes through their sidelobes. We build this property into the GP spectral-covariance model by decomposing the foreground covariance matrix into two separate parts,
\begin{equation}
K_{\mathrm{fg}} = K_{\mathrm{sky}} + K_{\mathrm{mix}}, 
\end{equation}
with `sky' denoting the intrinsic sky and `mix' denoting the mode-mixing contaminants. A Matern covariance function is adopted for each of the components of the GP model of the data, which is defined as~\citep{Stein99}:
\begin{equation}
\label{eq:matern_cov}
\kappa_{\mathrm{Matern}}(\nu_p, \nu_q) = \sigma^2 \frac{2^{1 - \eta}}{\Gamma(\eta)}\left(\frac{\sqrt{2\eta}r}{l}\right)^{\eta}K_{\eta}\left(\frac{\sqrt{2\eta}r}{l}\right),
\end{equation}
where $\sigma^2$ is the variance, $r = |\nu_q - \nu_p|$ is the absolute difference between the frequencies of two sub-bands, and $K_ {\eta}$ is the modified Bessel function of the second kind. The parameter $\eta$ controls the smoothness of the resulting function. Functions obtained with this class of kernels are at least $\eta$-times differentiable. The kernel is also parametrized by the hyper-parameter $l$, which is the characteristic scale over which the spectrum is coherent. Setting $\eta$ to $\infty$ yields a Gaussian covariance function, also known as the radial basis function, which is well-adapted to model the intrinsic (sky) foreground emission~\citep{Mertens18}. The coherence scale of this component is usually large, and we adopt a uniform prior $\mathcal{U}(10, 100)$ MHz for $l_{\mathrm{sky}}$. For the mode-mixing component, several covariance functions are evaluated. We test the Matern covariance function with different values of $\eta_\mathrm{mix}$ ($+\inf$, $5 / 2$ and $3 / 2$), and also the Rational Quadratic function ($\kappa_{\mathrm{RatQuad}}$) which was used recently in~\cite{Gehlot19a} to model the foreground contaminants of LOFAR-LBA data. A Matern kernel with $\eta_\mathrm{mix} = 3 / 2$ is favored by the data when comparing the Bayes factor (the ratio of the evidence of one hypothesis to the evidence of another), with very strong evidence against a wide range of alternatives (see Table~\ref{tab:all_gp_test} for a comparison of all tested GP models). A uniform prior $l_{\mathrm{mix}} \sim \mathcal{U}(1,10)$ is adopted, because simulations show that the foreground signal is separable from the 21-cm signal as long as $l_{\mathrm{mix}} \gtrsim 1$ MHz~\citep{Mertens18}.

\item[(2) The 21-cm signal ---]
The covariance shape of the real 21-cm signal is not known. However, information from current 21-cm simulations can be used to assess which family of models is a good approximation of the 21-cm signal.~\cite{Mertens18} show that the 21-cm signal frequency covariance -- calculated using 21cmFAST~\citep{Mesinger11} -- can be well-approximated by an exponential covariance function (i.e.\ a Matern function with $\eta = 1 / 2$). This function has two hyper-parameters: the frequency coherence scale $l_{21}$ and a variance $\sigma^2_{21}$. These allow some degree of freedom to match different phases of reionization. Based on the covariance of 21cmFAST simulations at different redshifts (see Figure 2 in~\citealt{Mertens18}), a uniform prior $\mathcal{U}(0.1,1.2)$ MHz on $l_{21}$ is adopted.

\begin{figure*}
    \includegraphics{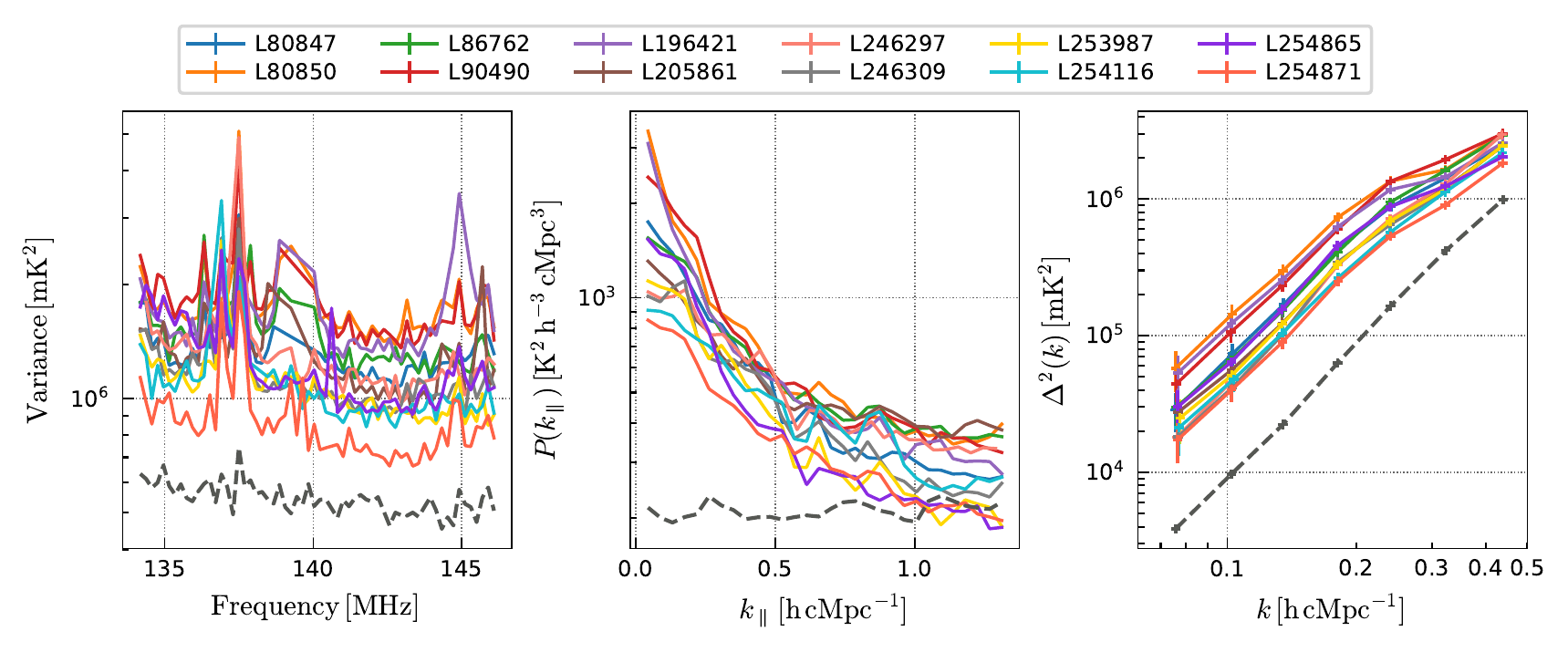}
    \caption{\label{fig:ps_all_nights_after_gpr}
    Variance (left panel), cylindrically averaged power spectra (averaged over all baselines) (middle panel) and spherically averaged power spectra (right panel) of Stokes I after GPR residual foreground removal, for all nights analysed in this work. The black dashed line represent the thermal noise power for an average observing duration time (14.4 h) and an average SEFD (4150 Jy). At high $k_{\parallel}$, the residual power after GPR is close to the thermal noise level, but a frequency correlated excess power is present. Note that the noise bias has not been removed here.}
\end{figure*}

\item[(3) The noise ---]
Various noise estimators can be used to build the noise covariance. The time-differenced visibilities -- obtained from the difference between even and odd sets of visibilities (e.g. separated by only several seconds) --  is expected to be an excellent estimator of the thermal noise. It does, however, not fully reflect the spectrally-uncorrelated random errors in our data (e.g.\ due to increased noise at short baselines; see Section~\ref{sec:noise_stats}). An alternative is to use Stokes V, which has previously been used as a noise estimator~\citep{Patil17}. It, however, can be corrupted by polarization leakage from Stokes I. The difference between alternating sub-bands in Stokes V can also be a good noise estimator, but it introduces correlation between consecutive sub-bands. The solution that is adopted, is to simulate the noise \lvekhidden{thermal?}covariance $K_{\mathrm{v_sn}}$ that we will use in our GP model using the weights in Eq.~\ref{eq:weights} and the noise definition of the gridded visibilities in Eq.~\ref{eq:sefd}. This estimator is based on Stokes-V noise, while the actual noise in Stokes I can be slightly higher (see Section~\ref{sec:noise_weights} and Table~\ref{tab:all_nights}). A noise scaling factor $\alpha_n$ is therefore adopted, which is optimized along with the other hyper-parameters of the GP model, resulting in the final noise covariance $\rm K'_{\mathrm{sn}} = \alpha_n K_{\mathrm{sn}}$. An associated noise data set $V_{\mathrm{N}}(u, v, \nu)$ is built to compute the noise power spectra and is used to subtract the noise bias from the residual power spectra. \lvek{I think this discussion is rather confusing still. There are many types of noises discussed and its not clear why the last one is adopted. I think this part needs a little more careful thought in how it is presented and why our chosen approach makes most sense.} \fms{I tried to rephrase a bit.}

\item[(4) The excess noise ---]
After applying GPR using foreground, 21-cm signal and noise-only covariance models, a significant spectrally-correlated residual is still present. This `excess noise or power' is accommodated in the model by an additional Matern covariance kernel $K_{\mathrm{ex}}$. Different values of $\eta_{\mathrm{ex}}$ were tested and  $\eta_{\mathrm{ex}} = 5 / 2$ is strongly favored by the data. Adding this `excess' component to the model significantly increases the Bayesian evidence (see Table~\ref{tab:all_gp_test}), motivating this choice.

\lvek{Refs to earlier work and papers in this.} \fms{Note that this is not the same excess as discussed in Patil et al. 2016.}

\end{description}

\begin{table}
\centering
\caption{Summary of the GP model, the priors on its hyper-parameters, and the estimated median and 68\% confidence intervals obtained using an MCMC procedure for the 10 nights dataset (see Appendix~\ref{sec:mcmc}. All covariance functions are Matern functions.}
\begin{threeparttable}
\label{tab:gp_model_fit}
\begin{tabularx}{\columnwidth}{XXX}
\toprule
Hyper-parameter & Prior & MCMC estimate (10 nights) \\
\midrule
$\eta_{\mathrm{sky}}$ & $+\infty$ & $-$ \\
$\sigma^2_{\mathrm{sky}} / \sigma^2_{\mathrm{n}}$ & $-$ & $611\substack{+22\\-19}$ \\
$l_{\mathrm{sky}}$ & $\mathcal{U}(10, 100)$ & $47.5\substack{+3.1\\-2.8}$ \\
\midrule
$\eta_{\mathrm{mix}}$ & $3 / 2$ & $-$ \\
$\sigma^2_{\mathrm{mix}} / \sigma^2_{\mathrm{n}}$ & $-$ & $50.4\substack{+2.1\\-1.9}$ \\
$l_{\mathrm{mix}}$ & $\mathcal{U}(1, 10)$ & $2.97\substack{+0.09\\-0.08}$ \\
\midrule
$\eta_{\mathrm{ex}}$ & $5 / 2$ & $-$ \\
$\sigma^2_{\mathrm{ex}} / \sigma^2_{\mathrm{n}}$ & $-$ & $2.18\substack{+0.09\\-0.14}$ \\
$l_{\mathrm{ex}}$ & $\mathcal{U}(0.2, 0.8)$ & $0.26\substack{+0.01\\-0.01}$ \\
\midrule
$\eta_{\mathrm{21}}$ & $1 / 2$ & $-$ \\
$\sigma^2_{\mathrm{21}} / \sigma^2_{\mathrm{n}}$ & $-$ & $ < 0.77$ \\
$l_{\mathrm{21}}$ & $\mathcal{U}(0.1, 1.2)$ & $ > 0.73^a$ \\
\midrule
$\alpha_n$ & $-$ & $1.17\substack{+0.06\\-0.06}$ \\
\midrule
\end{tabularx}
\begin{tablenotes}
  \small
  \item[a] The upper confidence interval hits the prior boundaries, hence we report here only the lower limit.
\end{tablenotes}
\end{threeparttable}
\end{table}

\noindent The final parametric GP model is composed of five terms:
\begin{equation}
  \rm K = K_{\mathrm{sky}} + K_{\mathrm{mix}}  + K_{\mathrm{21}} + K'_{\mathrm{sn}} + K_{\mathrm{ex}},
\end{equation}
with a total of nine hyper-parameters which we list in Table~\ref{tab:gp_model_fit}, along with their priors. An optimal GP model is obtained for each night separately by maximizing the Bayesian evidence. The Python package \texttt{GPy}\footnote{https://sheffieldml.github.io/GPy/} is used to do this optimization. The covariance parameters converge to very similar optimal values for all nights. The `sky' spectral-coherence scales are typically $l_{\mathrm{sky}} \sim 50 $ MHz, $l_{\mathrm{mix}} \approx 2.5 - 4.5$ MHz for the `mix' component and $l_{\mathrm{ex}} \approx 0.25 - 0.45$ MHz for the `excess' component. The `sky' component is expected to model emission from our Galaxy and extragalactic sources emitting predominately synchrotron and free-free radiation. These radiating sources have power-law spectra with temperature spectral-indices $\beta \sim 2.5$ for the Galactic synchrotron component~\citep[e.g.][]{Jelic08,Dowell17}, $\beta \sim 2.1$ for the free-free radiation~\citep[e.g.][]{Jelic08} and $\beta \sim 2.8$ for the extragalactic synchrotron component~\citep[e.g.][]{Lane14}. We verified experimentally that the coherence-scale $l_{\mathrm{sky}} \sim 50 $ MHz is well adapted to model power-law functions with spectral-index $\beta \approx 2 - 3$. The `mix' component is expected to model mode-mixing contaminants which in the cylindrically-averaged power spectra should be confined to the `foregrounds wedge' region. The coherence scale $l_{\mathrm{mix}} \approx 2.5$ of $K_{\mathrm{mix}}$ is associated with a step drop of power as function of $k_{\parallel}$, dropping to $\sim$1\% of the total power at $k_{\parallel} \approx 0.17 \mathrm{h\,cMpc^{-1}}$, and is thus well adapted to model this component. The variance of the `excess' is similar or below the noise variance ($\sigma_\mathrm{ex}^2 \approx 0.6 - 1\ \sigma_\mathrm{n}^2$) while for the `21-cm signal' it is typically very small ($\sigma_\mathrm{21}^2 < 0.1\ \sigma_\mathrm{n}^2$). Hence the residuals after removing the foregrounds are mainly composed of noise and `excess'. 

\lvekhidden{Again use of noise and power are often mixed. Also function and kernel, or term and component. Try to make the terminology uniform. I tried to fix some of this but not all.}


\begin{figure*}
    \includegraphics{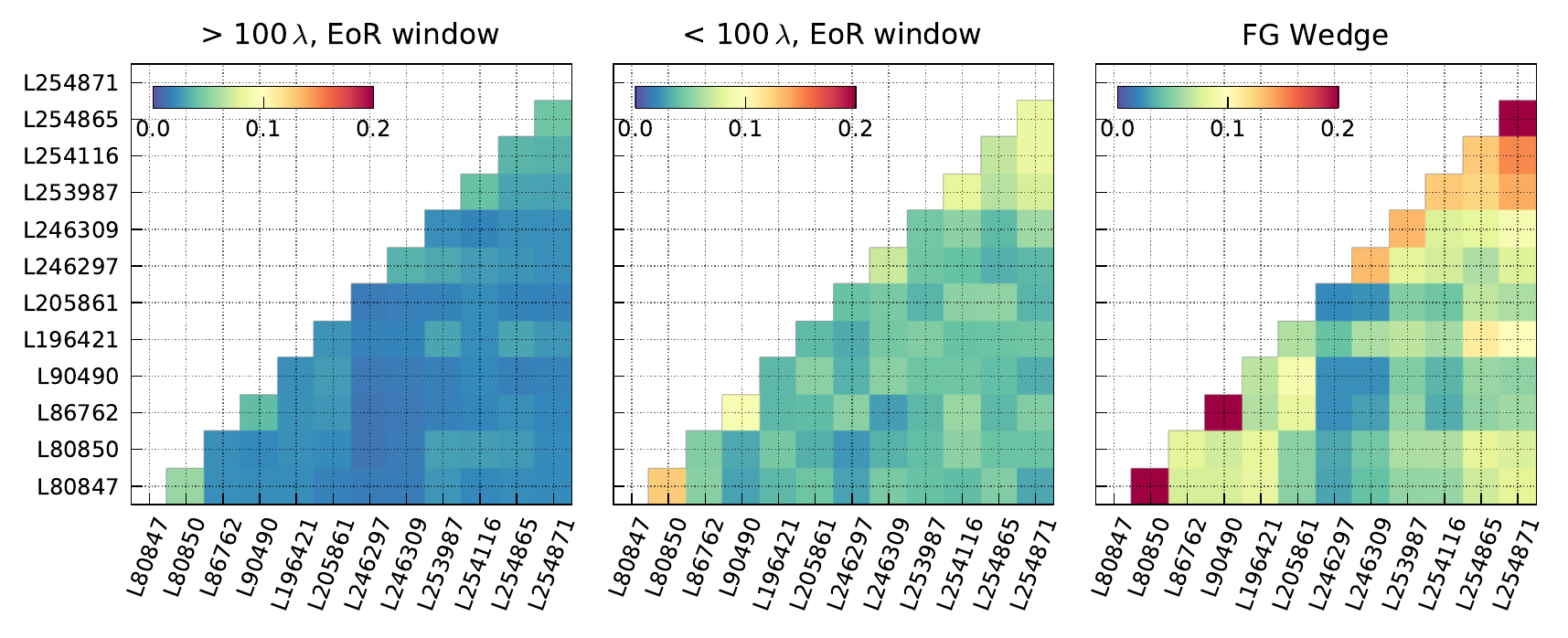}
    \includegraphics{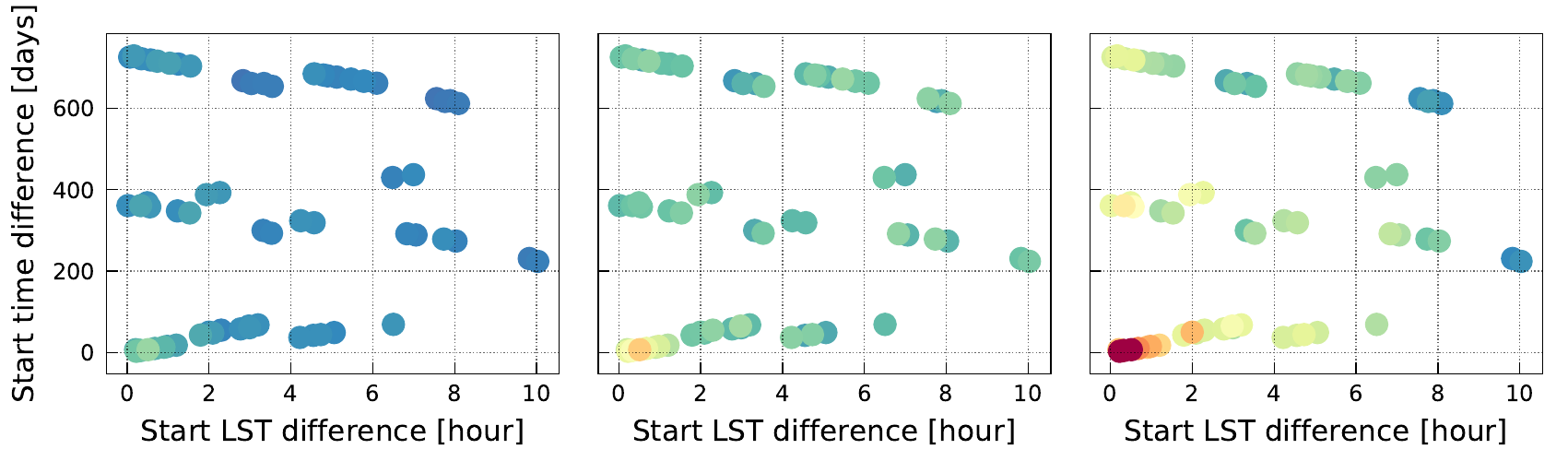}
    \caption{Top: \label{fig:nights_correlation_matrix}Cross-coherence matrix between
    all nights after GPR foregrounds removal. Three different regions of the
    cylindrically averaged power spectra are analysed: The EoR window for
    baselines $>100\,\lambda$ (left panel), the EoR window for baselines
    $<100\,\lambda$ (middle panel) and the foreground wedge region for
    baselines $>100\,\lambda$ (right panel). We note there is no or a small correlation in
    the EoR window, while the correlation is more noticeable in the foreground
    wedge, especially for certain combinations of nights. Bottom: \label{fig:nights_correlation_lst_obs_time}Cross-coherence (color scale)
    between two nights as a function of LST time difference (abscissa) and UTC
    time difference (ordinate). We observe
    higher correlation between observation started at the same LST time (which
    will see the \textit{same} sky throughout the observation). \lvekhidden{Add title above plots just as in Fig.6. Or merger Fig 6 and 7, since also this figure has no color wedge. When printed the two figures might not be next to each other and hence having no title/color-wedge might complicate the interpretation.}}
\end{figure*}
\vspace{1cm}

\subsubsection{Power spectra after foreground removal}

Figure~\ref{fig:ps_all_nights_after_gpr} shows the variance and power spectra of the residual after GPR foreground removal for all nights, compared to the expected thermal noise level for an average observing duration time of 14.4\,h with an SEFD of 4150\,Jy. For all nights, the excess power per sub-band is a factor of $2 - 3$ times higher than the thermal noise. This excess corresponds to the `excess' component of our GP model which is not removed from the data due to its small frequency coherence scale. At small $k_{\parallel}$, the ratio of residual to thermal noise power is $\approx 5 - 10$, while it is $\approx 1 - 2$ at large $k_{\parallel}$. The same can be seen in the spherically averaged power spectra. Night-to-night variations of the residual power is a factor $2 - 3$ and cannot be explained by the different total observing times between nights. For example, the excess power in LOFAR observing-cycle 2 observations is below that for cycles 0 and 1. Different ionospheric or RFI conditions might contribute to these night-to-night variations. Hence, although this excess power is drastically lower than in~\cite{Patil17} due to improved calibration, it is still not entirely mitigated. Below we investigate the excess power in more detail.


\begin{figure*}
    \includegraphics{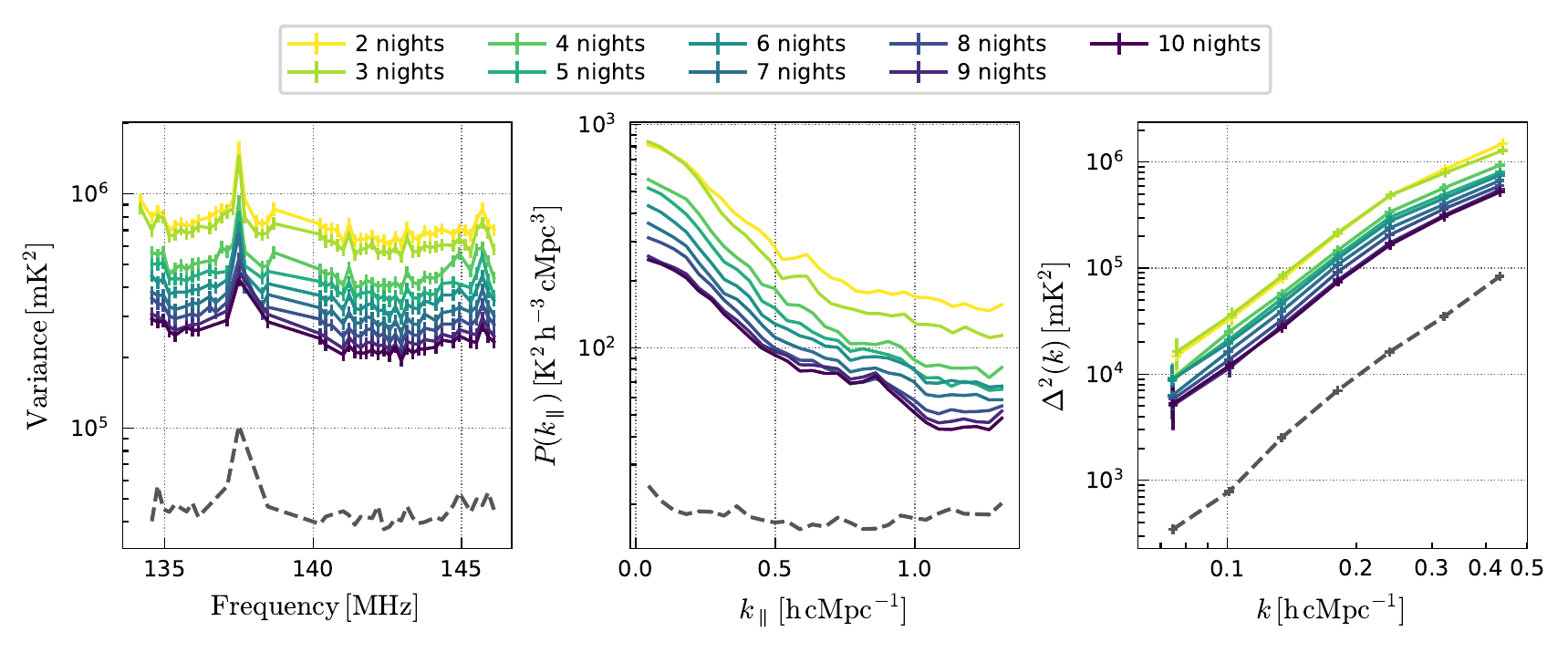}
    \caption{\label{fig:ps_all_combined_after_gpr}
    Variance (left panel), cylindrically averaged power spectra (averaged over
    all baselines) (middle panel) and spherically averaged power spectra (right
    panel) of Stokes I after GPR residual foreground removal, as we combine the
    nights, from 2 (yellow) to 10 (dark blue). The black dashed line
    represent the thermal noise power of the 10 nights dataset, estimated from
    10\,s time difference visibilities. Note that the noise bias is not removed
    here.}
\end{figure*}

\subsection{Night-to-night correlations between residuals}
\label{sec:night_to_night_correlation}

To better understand the origin of the excess power after foreground removal, the residuals obtained after GPR foreground removal are correlated between all pairs of nights, by computing the cylindrically-averaged cross-coherence, defined as:
\begin{equation}
  C_{1,2}(k_{\perp}, k_{\parallel}) 
\equiv \frac{ \left< |\tilde{T^*_1}(\mathbf{k}) \tilde{T_2}(\mathbf{k}) |
  \right>^2}{\left< |\tilde{T_1}(\mathbf{k})|^2 \right> \left< |\tilde{T_2}(
  \mathbf{k})|^2 \right>},
\end{equation}
\lvekhidden{The square is outside the average in the numerator, which I think is incorrect. Double check this.} which is a normalized quantity between one (indicating maximum correlation) and zero (no correlation). The cylindrically-averaged cross coherence is computed between all pairs of nights. The average over three regions in $(k_{\perp}, k_{\parallel})$ space is determined: the ``foregrounds wedge'' region bounded by the instrumental horizon delay line (see Eq.~\ref{eq:wedge})\lvekhidden{Wedge in the usual sense, i.e. increase in k-parallel with k-perp? Above horizon line? Explain clearer.} and two EoR-window regions distinguishing between the shorter ($|\mathbf{u}| < 100$; roughly the central LOFAR `superterp' region) and the longer core-baselines. 
This allows an additional test of whether the night-to-night correlations of the excess noise described in Section~\ref{sec:noise_stats} correlate with where it is found in the power spectrum and correlates with baseline length.

\lvekhidden{Again above horizon-delay line?}

A corner-plot of the correlations between nights is presented in Figure~\ref{fig:nights_correlation_matrix} for each of the three different regions. We also show the correlation coefficients as a function of their difference in the start of the observations in Local Siderial Time (LST) versus their start in number of (Julian) days. This representation provides additional clues about the different observing conditions between nights. In the `EoR window', only very small correlations are observed. The correlation is on average slightly larger for the shorter baselines ($\approx$0.04, significance $> 0.032$) than for the larger baselines ($\approx$0.02, significance $> 0.018$), as defined above. Significantly larger correlations are found in the ``foregrounds wedge'' region ($\approx 0.03 - 0.25$, significance $> 0.018$). For each of the three regions, also a clear trend between the correlation coefficients and either difference in Julian date (between nights) or LST are found: correlations are larger if the observations are either close in Julian date or close in LST, and largest if they are close in both, hence they observe the same sky during the observing runs with a similar primary beam and a similar PSF. The largest correlation, in particular inside the wedge region, is found when two nights are close and separated by only a small number of days. This suggests that some of the excess power in the data residuals (after sky-model and foreground subtraction) originates from sky emission that is far from the phase center for which the primary beam will change considerably at different values of the LST. The PSF will also change but, for all nights, the uv-plane is always fully sampled in the $50 - 250 \lambda$ range, given the long (12h to 16h) duration of our observing nights. For the shorter baselines and in the `EoR window' region, the trend with LST difference is less pronounced, which suggests that part of the additional noise at baselines $< 100 \lambda$ discussed in Section~\ref{sec:noise_stats} may have a local origin (e.g. RFI). These are all baselines from stations in the superterp and might arise from mutual-coupling. Its origin will be investigated in the future using a near-field imaging technique~\citep{Paciga11}.

\lvekhidden{Local to what? The superterp? Why would that care about LST or UT? Local RFI would move around with the Earth and hence not be affected by LST, right?}

\lvekhidden{Such as? Maybe add some comments.}
\lvekhidden{Can you provide the correlation coefficient for a night with only uncorrelated noise at the level of the excess?}
\lvekhidden{Give range of values for the latter correlations.}
\lvekhidden{What about PSF?}
\lvekhidden{Longer?}

Based on this analysis, we discard nights L80850 and L254871 as the former has a high residual power and both have a high correlation coefficient between their residuals with other nights. This leaves a total of ten nights for further analysis.

\lvekhidden{The description in this section was confusing. I tried to clarify this, but please check if I interpreted this correctly.}

\section{Combining data sets}
\label{sec:results_combined}

In this section, we discuss the power spectra  obtained by combining the ten selected nights of observations, corresponding to about 141 hours of
data.

\begin{figure*}
    \includegraphics{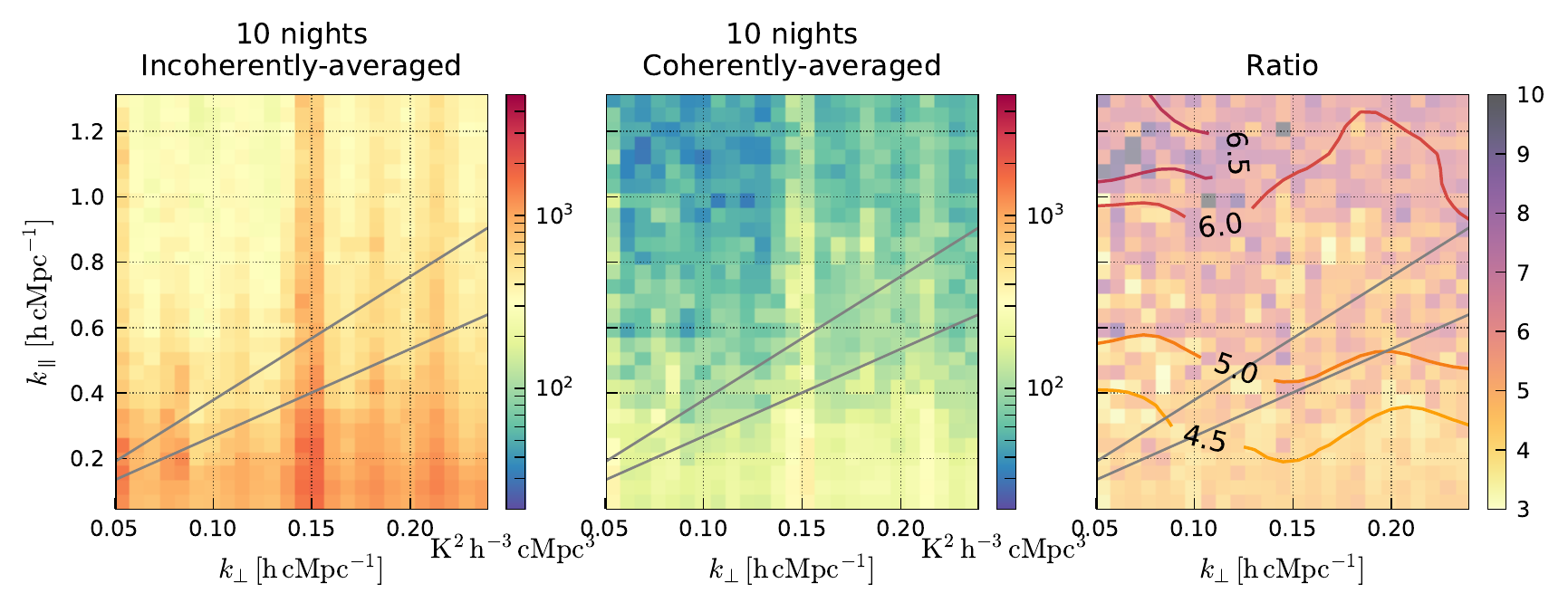}
    \caption{\label{fig:ps2d_1_vs_10_nights}
    Cylindrical Stokes I power spectra after GPR residual foreground removal
    of the 10 nights incoherently averaged (left panel) and coherently averaged
    (middle panel). Both are optimally combined and thus the ratio of the two
    (right panel) is expected to be 10 in the case of uncorrelated residuals. We
    observe significant residuals in the foreground wedge region, especially
    below the $50 \degree$ delay line (black lines), for both the incoherently and
    coherently averaged cases. The ratio of the two is $<5$ in this region,
    suggesting frequency correlated excess power which is also partially correlated between nights.}
\end{figure*}

\begin{figure*}
    \includegraphics{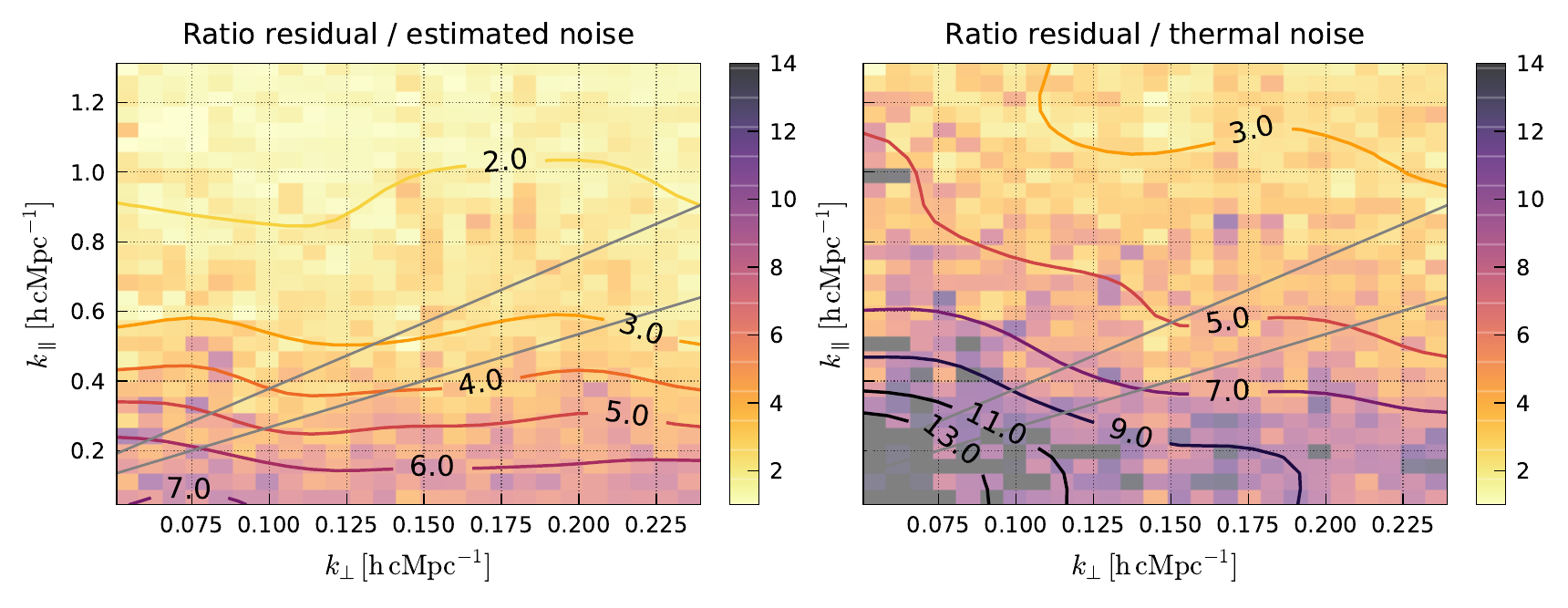}
    \caption{\label{fig:ps2d_residual_over_noise}
    Ratio of cylindrical Stokes I power spectra of the 10 nights Stokes I after
    GPR residual foreground removal over the noise estimated by GPR (left
    panel) and the thermal noise estimated from 10\,s time difference
    visibilities (right panel). The excess power (against the frequency
    uncorrelated noise) does not show strong baselines dependence. The baseline
    dependence of the excess noise (described in Section~\ref{sec:noise_stats})
    is striking when compared against the thermal noise. \lvekhidden{it might be useful to add the estimated noise over thermal noise ratio as well for completeness.}}
\end{figure*}

\subsection{Weighted averaging of the  data}

The gridded visibilities of separate nightly data sets are averaged following the procedure described in Section~\ref{sec:nights_averaging}. They
are combined in the order of their date of observation\footnote{This is only done for illustration purposes, since the final result does not depend on the order in which the data is combined.}. Intermediate data sets are also kept, yielding a total of nine combined data sets with an increasing total observation time. For each accumulated data set, the residual foregrounds are estimated and subtracted following the same GPR procedure and GP covariance model described in Section~\ref{sec:gpr_fg_removal}. Hence, the GPR  is only applied to the combined data sets.

\lvekhidden{I rephrased the paragraph below since it was hard to understand. Double check if still ok.} 

When combining the data, the GP spectral coherence scales of the foregrounds converge to similar values as found from individual nights. This suggests that these scales are stable between nights. The GP variances for the `sky' and `mix' components also do not vary much when compared to the total variance ($\approx 0.85 - 0.9$ for the `sky' component, and $\approx 0.04 - 0.065$ for the `mix' component). This is expected for a signal that is coherent over nights. The GP variance of the `excess' component decreases with increasing total observation time. It does not decrease, however, as would be expected from uncorrelated noise, with a ratio $\approx\,2.2$ found between the two nights data set (28\,h) and ten nights data set (141\,h), confirming that the `excess' component partly correlates between nights. The most probable hyper-parameter values for the combined (i.e.\ ten nights) data set are given in Table~\ref{tab:gp_model_fit}, with their confidence intervals obtained using an MCMC procedure (see Appendix~\ref{sec:mcmc} and~\citealt{Mertens18}). Most parameters are well constrained, except the variance of the `21-cm signal' component which is consistent with zero, as expected for such a short total integration time, and the coherence-scale of the `21-cm signal' for which the upper bound of the posterior distribution hit the prior boundary, also the significance of the later is reduced given the non-significant variance of this component. Hence only upper limits on the 21-cm signal (power spectra) can be given. 

\lvekhidden{How much? consistent in the errors? Make this clear.}
\lvekhidden{You mean the actually excess power or variance of the power?}
\lvekhidden{Give which ratio for 10 nights}

\subsection{Residual power spectra}

Figure~\ref{fig:ps_all_combined_after_gpr} shows the power spectrum and its integrated variance after applying GPR, but {\sl before} subtracting the noise bias, as we combine more data. The frequency range of $136 - 140$ MHz is heavily affected by RFI and many of the corresponding sub-bands are therefore discarded. The results are compared to the thermal noise power estimated from the 10\,s time difference visibilities. The data are combined (i.e. integrated) in the order of the observation. The integrated variance as a function of frequency (left panel) shows a gradual reduction of power as we combine more data. However, taking the ratio between the 2 and 10 nights of accumulated data, a value of $\approx$3 is found while theoretically a ratio closer to $\approx$5 is expected. Examining the power spectra as a function of $k_{\parallel}$ (middle panel) shows that the ratio of residual power over thermal noise is worse in the foreground-dominated region (i.e. inside the `wedge'), where only a reduction in power of $\approx$2.8 is found. At $k_{\parallel} > 1 \mathrm{h\,cMpc^{-1}}$, the ratio is closer to $\approx$4. Comparing the residual power to the thermal-noise power in the spherically averaged power spectrum (right panel), the residual power is found to be $\approx$14 times the thermal noise power at $k \approx 0.08 \mathrm{h\,cMpc^{-1}}$, and about $\approx$6 times the thermal noise power at $k \approx 0.45\mathrm{h\,cMpc^{-1}}$. 

In Figure~\ref{fig:ps2d_1_vs_10_nights}, we compare the cylindrically-averaged power spectra of the 10 nights data set residual (middle panel) to a 1 night equivalent data set power spectrum in which the different nights are averaged incoherently (i.e. averaged in power spectra) (left panel). Taking the ratio of the two (right panel), we observe a ratio $\approx$4 in the foreground wedge region and $\approx 5-6$ outside it where a ratio of 10 is expected. This indicates that the night-to-night correlation of the residual is not just limited to the wedge, where some residual sky foregrounds might be expected, but also affects the EoR window. Even at high $k_{\parallel}$, the residuals are not thermal noise dominated in the combined data set. This night-to-night correlation of the residuals, that we also observed in Section~\ref{sec:night_to_night_correlation}, is the major challenges that needs to be understood and solved in the future as it limits our ability to integrate >200 hours of data. Possible origins will be discussed in Section~\ref{sec:discussion}.

\lvekhidden{This needs a clear statement, and not left dangling. This is clearly a huge issue. Better address is on the spot and reference a later discussion as well.} \fms{I tried to rephrase the above, hopefully it reads better.}


\subsubsection{Residual over thermal-noise power ratio}

Figure~\ref{fig:ps2d_residual_over_noise} shows the ratio of the power spectrum of the Stokes-I residuals over the observed noise power spectrum (left panel) and over the thermal noise power spectrum (right panel). The noise power spectrum is computed from the simulated noise data set $V_{\mathrm{N}}(u, v, \nu)$ used in the GP model (see Section~\ref{sec:cov_model}) and accounts for the larger spectrally-uncorrelated noise level observed on baseline lengths of $< 125\,\lambda$ as compared to the thermal noise. Hence, it incorporates the noise scaling factor $\alpha_n$ which is optimized as part of the GP covariance model. The residual of the Stokes-I over the observed noise ratio shows that the GP model properly accounts for the spectrally-uncorrelated noise in the data: a ratio $\sim 1$ is reached at $k_{\parallel} > 1 \mathrm{h\,cMpc^{-1}}$. At lower values of $k_{\parallel}$, however, the ratio gradually increases. This is the spectrally-correlated excess power, which is also part of the GP model, but is not part of the foreground covariance model. Remarkably, the ratio appears to be baseline independent, indicating that the excess power follows the same baseline dependence as the noise (which corresponds to the uv-density). Examining the ratio of the residual over the thermal noise shows that it increases toward shorter baseline lengths.

In summary, the residual power spectrum from the combined data set, after GPR foreground removal, can be decomposed into (i) thermal noise, (ii) an additional noise-like component that is spectrally uncorrelated, and (iii) an excess noise that is partially correlated between nights and spectrally-correlated (i.e. its power spectrum in delay space is not white) and cannot be removed by the GPR method as part of the spectrally-smooth foregrounds. The noise power is still significantly larger than the thermal noise power, especially on shorter baseline lengths, although the excess is much smaller than found in~\citet{Patil17} due to the signal-processing improvements presented in this paper.

\subsection{Upper limit on the 21-cm signal power spectrum}

\begin{figure}
    \includegraphics{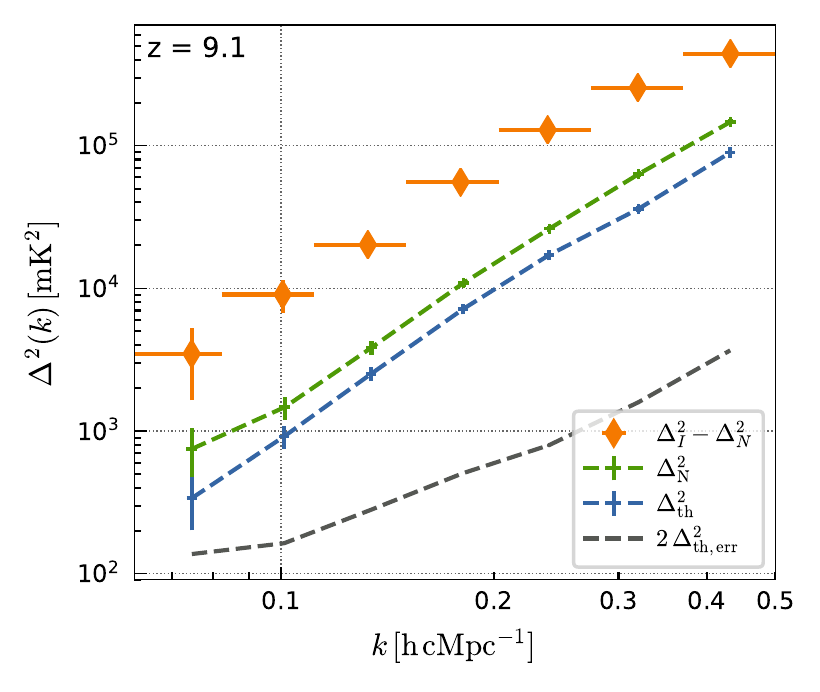}
    \caption{\label{fig:ps_final_result}
    Final 10 nights Stokes I spherically averaged power spectra after GPR
    residual foreground removal and noise bias removal (orange). The green and
    blue dashed lines represent, respectively, the estimated frequency-uncorrelated 
    noise and thermal noise power of the 10 nights dataset. The
    black dashed line represents the $2-\sigma$ upper limit theoretically achievable
    if the residual of the 10 nights dataset were thermal noise dominated.}
\end{figure}

The spherically-averaged power spectrum is
computed inside seven $k$-bins logarithmically spaced between $k_\mathrm{min} = 0.06
\mathrm{h\,cMpc^{-1}}$ and $k_\mathrm{max} = 0.5 \mathrm{h\,cMpc^{-1}}$, with
a bin size of $\mathrm{d}k / k
\approx 0.3$. Assuming that (a) the GPR foregrounds have limited impact on the power spectra of the 21-cm signal (see Appendix~\ref{sec:inj_simu_test}), and that (b) the power spectra of the noise $V_{\mathrm{N}}(u, v, \nu)$, estimated as part of the GP covariance model optimization, are a good representation of the spectrally-uncorrelated noise power in our data set, we can compute the spherically-averaged noise subtracted power spectrum of the residual and its associated error as:
\begin{align}
&\Delta^2_{21} = \Delta^2_{I} - \Delta^2_{N} \\ 
&\Delta^2_{21,\mathrm{err}} = \sqrt{\left(\Delta^2_{I,\mathrm{err}}\right)^2 + \left(\Delta^2_{N,\mathrm{err}}\right)^2}.
\end{align}
%
The resulting power spectrum is presented in Figure~\ref{fig:ps_final_result}. It significantly exceeds both the thermal noise power $\Delta^2_{\mathrm{th}}$ and the estimated noise power $\Delta^2_{N}$, because on large scales it is dominated by the excess power described in previous sections. Although the value of $\Delta^2_{21}$ for the combined data sets is significantly larger than zero, we do not consider it a detection. The reason is that the residuals are only partially correlated between nights whereas the 21-cm signal would be fully correlated (assuming it dominates the noise), and it is not isotropic (i.e.\ constant power for all modes of a given $k$). Conservatively, we therefore consider it to be an upper limit on the 21-cm signal and report the $2-\sigma$ upper limits in Table~\ref{tab:upper_limit}.

The deepest upper limit $\Delta^2_{21} < (72.86)^2\,\mathrm{mK^2}$, is observed at $k = 0.075 \mathrm{h\,cMpc^{-1}}$. Despite it being the deepest upper limit at this redshift, this is still a factor $\sim30$ higher in power than the upper limit that could theoretically be achieved if the residual would be consistent with thermal noise. To make a comparison with the previous upper limits based on 13\,h of data~\citep{Patil17}, we  note that in the present work we discard the smallest $k_{\parallel}$ modes when computing the spherically averaged power spectra while this was not the case in~\cite{Patil17}, limiting the smallest measurable $k$ mode\footnote{The smallest $k$ bin in~\citealt{Patil17} was $0.053 \mathrm{h\,cMpc^{-1}}$.}. We also use different foregrounds-removal and power spectrum estimation methods. Nevertheless, at $k = 0.1 \mathrm{h\,cMpc^{-1}}$, the upper limit on $\Delta^2_{21}$ is improved by a factor 7.7. \lvekhidden{Make explicit whether this is also for 13\,h or for 141\,h of data, this comparison.} Most of this improvement can be attributed to the improved DD calibration.

\begin{table}
\centering
\caption{$\Delta_{21}^2$ upper limit at the 2-$\sigma$ level ($\Delta_{21,\mathrm{UL}}^2$) and theoretical thermal noise sensitivity ($\Delta_{\mathrm{th,err}}^2$) from the 10 nights data set, at given $k$ bins. \lvekhidden{Explain the last column better. It's a bit odd now. Also since we use $\delta k \ll k$ it's not easy to compare to other papers. Our limits could be deeper if we average over a larger bin-size, using the Golden rule.}}
\begin{threeparttable}
\label{tab:upper_limit}
\begin{tabular}{rrrrr}
\toprule
$k$ &  $\Delta_{21}^2$ &  $\Delta_{21,\mathrm{err}}^2$ & $\Delta_{21,\mathrm{UL}}^2$ & $2\,\Delta_{\mathrm{th,err}}^2$
\\
$\mathrm{h\,cMpc^{-1}}$ & $\mathrm{mK^2}$ & $\mathrm{mK^2}$ & $\mathrm{mK^2}$ & $\mathrm{mK^2}$ \\
\midrule
0.075 & $(58.96)^2$ & $(30.26)^2$ & $(72.86)^2$ & $(13.10)^2$ \\
0.100 & $(95.21)^2$ & $(33.98)^2$ & $(106.65)^2$ & $(14.30)^2$ \\
0.133 & $(142.17)^2$ & $(39.98)^2$ & $(153.00)^2$ & $(18.73)^2$ \\
0.179 & $(235.80)^2$ & $(51.81)^2$ & $(246.92)^2$ & $(25.16)^2$ \\
0.238 & $(358.95)^2$ & $(64.00)^2$ & $(370.18)^2$ & $(31.54)^2$ \\
0.319 & $(505.26)^2$ & $(87.90)^2$ & $(520.33)^2$ & $(44.60)^2$ \\
0.432 & $(664.23)^2$ & $(113.04)^2$ & $(683.20)^2$ & $(67.76)^2$ \\
\midrule
\end{tabular}
\end{threeparttable}
\end{table}

\section{Discussion}
\label{sec:discussion}

In this section, a number of checks of the results of our processing pipeline are discussed. Further improvements to the upper limit by investigating potential sources for the still large excess power and mitigation methods are also discussed.

\subsection{Data-processing cross-checks}
\label{sec:cross-check}

A critical assessment of the full processing pipeline is essential to ensure a reliable upper limit on the 21-cm signal. Such a complex experiment uses advanced signal processing techniques that may potentially remove or alter the signal if not applied properly (and sometimes even if they are applied properly). \lvekhidden{Cite the PAPER papers as a case study where this went wrong already here?} A number of such scenarios have been documented as a result of biases in the calibration~\citep[e.g.][]{Patil16,Barry16,Ewall17}, foregrounds mitigation~\citep[e.g.][]{Paciga13} and power spectra estimation~\citep[e.g.][]{Cheng18,Kolopanis19}. To ensure limited signal loss or bias of the 21-cm signal power spectra, a number of checks were performed at various steps in the processing pipeline.

\lvekhidden{New PAPER papers can now be added.} 

\begin{description}
    \setlength\itemsep{1em}

    \item[Calibration ---] Direction-dependent calibration has the potential to modify the signal when solving for too many parameters~\citep{Patil16}. Our calibration scheme strictly limits this possibility by discarding the baselines $<\,250\,\lambda$ during the calibration step and enforcing spectral smoothness of the instrumental gains via regularisation. This bias reduction was  also verified theoretically~\citep{Sardarabadi19} and experimentally (Mevius~et~al.~in~prep.). We additionally checked that the Stokes-V power spectra before and after DD-calibration are comparable, checked that images of Stokes Q and Stokes U show the same diffuse Galactic polarized structure before and after DD-calibration (only point sources due to polarization leakage are removed, as expected), and checked that we observe the same polarized structure at Faraday depths of $-30$ and $-24.5 \,\mathrm{rad\,m^{-2}}$ , before and after DD-calibration, as previously observed in~\citet[][Figure 3]{Patil16}. In each of the cases, we confirm that diffuse emission is not suppressed on baselines $<\,250\,\lambda$ where we determine the 21-cm signal results, as expected since they do not participate in the calibration. 

    \item[Foregrounds mitigation ---] The GPR foregrounds mitigation method has been extensively tested against a large range of foreground simulations~\citep{Mertens18} as well as simulated LOFAR~\citep{Offringa19a} and SKA foregrounds~(Mitra~et~al.~in prep.). \cite{Mertens18} showed that statistical separation between foregrounds and signal can be achieved when the foregrounds are correlated on frequency scales $\gtrsim 3$ MHz which is the case for the combined data set ($l_{\mathrm{mix}} = 3.0 \pm 0.1$ MHz). We can also recover an unbiased power spectrum of the signal when the chosen GP covariance model is a good match to the data. In reality, the model and data might not be perfect matches, and some biases can be expected. To assess this, injection tests and simulation tests were performed which reproduce the frequency correlations in the data. The results are presented in Appendix~\ref{sec:inj_simu_test} and Figures~\ref{fig:gpr_injection_test} and~\ref{fig:gpr_simulation_test}. No signs of significant signal loss are found in any of tested cases. The 21-cm signal is recovered effectively unbiased in the simulation tests. In the injection test, we observe a positive bias $< 3$ on large scales and low S/N which is reduced to $\sim$1 at higher S/N scenario.\lvekhidden{What scales and what s/n? Moreover, is a factor of 3(!) really a limited bias? I would say no. Also make sure you mention whether we correct for this bias or not.}

    \item[Power spectra ---] The power spectra estimation has been tested against a data set with known power spectra as part of a SKA blind challenge (Mitra et al. in prep.) and has been compared to other power spectra pipelines~\citep[e.g.][]{Offringa19a} demonstrating the accuracy of our power spectra pipeline. Uncertainty estimates are tested using a Monte Carlo method with noise and simulated 21-cm like signals showing good agreement between our analytical estimates and the ones obtained from simulations.

\lvekhidden{Should we add this to the appendix? Otherwise it's unlikely to ever be published.}

\end{description}

\lvekhidden{You don't mention the conclusion of these tests in some cases. Be clear on this, e.g. in case (3). Maybe write an overall summary here of these tests and how we deal with observed biases.}

\begin{figure*}
    \includegraphics{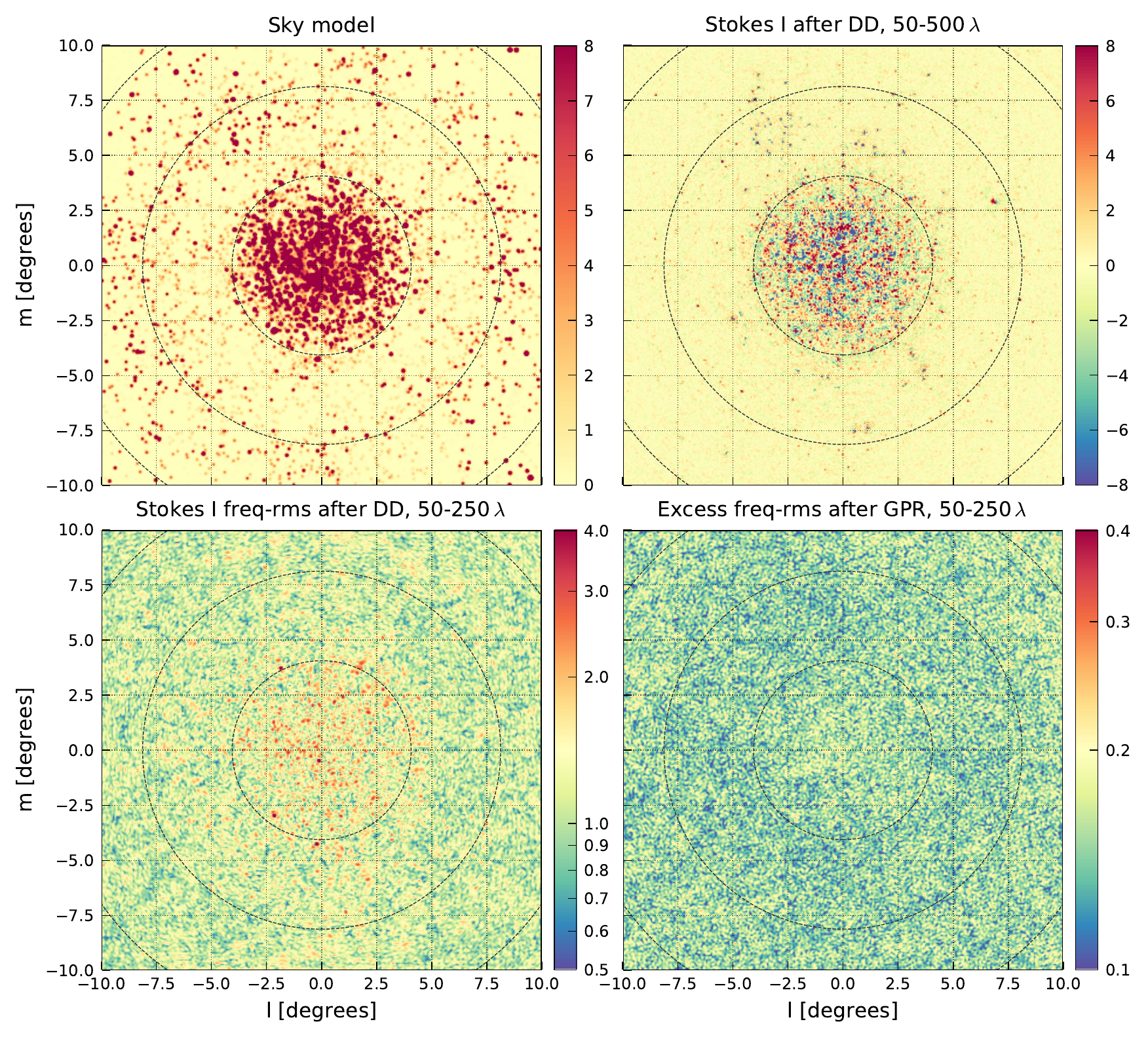}
    \caption{\label{fig:images_sky_model+after_dd+after_gpr}
    Top-left: Apparent NCP sky-model, convolved with a 7 arcmin FWHM Gaussian
    PSF, composed of more than 28000 components distributed in 122 clusters.
    Top-right: 10 nights total intensity (Stokes I) image averaged over the 12
    MHz bandwidth after DD-calibration and sky-model subtraction, at 7 arcmin
    resolution. Bottom-left: 10 nights total intensity image rms along the
    frequency-direction, after DD-calibration and sky-model subtraction, at 13
    arcmin resolution. Bottom-right: 10 nights total intensity image rms along
    the frequency-direction after GPR residual foreground removal, at 13 arcmin
    resolution. All images are in units of Kelvin, and the three dashed circles
    indicate the approximate position of the primary beam nulls ($\approx$
    4.5, 9, 13.5 degrees).}
\end{figure*}

\subsection{Possible origin of the excess power}
\label{sec:origin_excess_power}

The residual power spectra after GPR foreground removal and noise bias subtraction are dominated by an excess power that is in part spectrally and temporally (i.e.\ between nights) correlated. On large angular scales ($k \approx 0.1 \mathrm{h\,cMpc^{-1}}$), this excess power reaches $\approx 22$ times the thermal-noise power (Fig.~\ref{fig:ps_final_result}), and currently it is the dominant effect that impacts our 21-cm signal upper limits (or its future detection) with LOFAR. In the ideal situation where one is thermal noise limited, by combining >100 nights of data (roughly the data in hand), limits of a few mK at $k = 0.1 \mathrm{h\,cMpc^{-1}}$ can in principle be reached. Understanding the origin of this excess power is therefore essential. Below, we discuss several potential causes. A more detailed analysis is left for a forthcoming work (Gan~et~al.,~in prep.).

\lvekhidden{How large? Also indicate figure nr. again.}
\lvek{Mention that this is in the realm of 21-cm signal models, but of course there is a wide range of models, some fainter from brighter. We need a small discussion or statement here on whether this limit is interesting and why.}

\begin{description}
  \setlength\itemsep{1em}
  
  \item[Foreground sources ---] Most of the foreground sources and their associated PSF side-lobes are subtracted during DD-calibration and the GPR foreground-removal steps. In Figure~\ref{fig:images_sky_model+after_dd+after_gpr}, a $20\degree \times 20\degree$ image of the sky model is presented, restored with a 7-arcmin FWHM Gaussian PSF (top-left panel) as well as an image of the frequency-averaged (continuum) Stokes-I image after DD-calibration (top-right panel). Most of the sources from the sky model are correctly subtracted. The main lobe of the Primary Beam (PB) is confusion noise limited on this angular scale and dominates the residual foregrounds. The standard deviation in the frequency direction of the DD-calibrated image cube (bottom-left panel), indicates that although most of the line-of-sight power is inside the main PB lobe, there is significant power outside as well. After GPR (bottom-right panel), the residual power becomes more spread over the full field but remains concentrated mainly inside the first and second null of the PB. There is no significant correlation between (a) the variance in the frequency direction after GPR and (b) the structure in the Stokes-I image after DD-calibration or the sky-model image. This suggests that (i) GPR properly removed the confusion limited foregrounds in the inner $<\,20\degree$ from the phase center, and (ii) the excess power does not originate predominately from sources $<\,20\degree$ from the phase center. The larger coherence found between two nights observed at similar LST ranges and the decorrelation at larger LST time difference (Fig.~\ref{fig:nights_correlation_matrix}) could also be explained by this hypothesis given that the average PB only changes significantly between LSTs at distance $>\,20\degree$. Foreground sources further from the beam center that are not part of the sky model result in spectrally fluctuating side-lobes, due to the chromatic PSF, that GPR might find hard to model. The Galactic plane, which is about $30 \degree$ from the NCP, is very bright on large spatial scales and could also be a source of the excess power. However, in the cylindrically averaged power spectra, its power should still be limited to the foreground wedge, while this is not the case for the excess (Fig.~\ref{fig:ps2d_residual_over_noise}) which has power up to high $k_{\parallel}$ and no clear baseline dependence.

  \lvek{Why? That's not self-evident. This needs a better explanation. Do you mean that it's not the gain errors affecting the sky-model? Would gain errors applied to the sky model not reflect the primary beam on Fig.12 (bottom right) since that contains most power? How can it be independent from the sky model or where the power is in the sky? } \fms{Only considering the foregrounds sources, no side-lobes, no gains errors, if you had left foregrounds sources in the $<\,20$ degrees, you would expect some correlation before/after DD.}
  \lvek{Why? I would conclude the opposite.} \fms{I tried to rephrase a bit.}
  \lvek{I find it hard to believe that the PB is the problem. It has at most internal baselines of 16 lambda, and hence any PB effect has very small delays. The effect comes from a combination of the different PB between stations on longer baselines.} \fms{Not PB side-lobes, but PSF side-lobes.}
  \lvekhidden{Ref to figure}
  
  \item[Polarization leakage ---] LOFAR has an instrumentally-polarized response.  This may cause diffuse polarized emission to leak into Stokes I. Faraday rotation of the polarized foreground could then introduce spectral fluctuations, which may mimic or cover up the frequency structure of the 21-cm
  signal~\citep{Jelic15,Asad15}. Although this could explain the spectral correlation of
  the excess power, the predicted level of leakage is expected to be much smaller (i.e.\ $\sim 1 \%$) than the observed level of excess power \citep[see][]{Asad16}. Hence, we believe that the current level of excess power is not the result of polarization leakage in the NCP, which is only marginally polarized.

  \item[DD-calibration errors ---] The over-fitting of the data in the DD-calibration step caused by the removal of baselines $<250\,\lambda$ during calibration in the past has been a clear origin of excess power in LOFAR data (see the discussion in e.g.~\citealt{Patil16, Patil17} and the simulation from~\citealt{Sardarabadi19}). The improvements made
  in the calibration step have considerably reduced its impact, and it should not introduce the kind of excess we observe in the full 141\,h data set. To verify this, the  power spectrum of the DD-calibrated sky model for one night (i.e.\ L253987) is created, showing negligible power above the wedge. We therefore conclude that the DD-calibrated sky-model in our current approach  is sufficiently spectrally smooth that it does not leak power in the EoR window. On the other hand, no DD-calibration is applied to the residuals after sky-model subtraction (e.g.\ confusion-level sources and diffuse emission that are not part of the sky-model) which only have DI-calibration gain applied to them.

  \lvekhidden{Reference/mention again the paper by Millad proving this is indeed the case.}
  \lvekhidden{How do we know this for the current data, which has not been processed optimally yet? Could the current excess still be the same excess but at a lower level?}
  \lvek{Can we add this figure? It would help a lot. Can we argue that this power is also below the expected 21-cm signal? Also showing a ration of the model power and the excess power or thermal noise would tell us where the model is sub-dominant.} \fms{I'll think about it.}
  \lvek{Can we also reference Barry et al. here?} \fms{I don't think it fit here, they do not do DD-calibration.}

  \item[DI-calibration errors ---] At present, the spectral smoothness via Bernstein polynomials in \texttt{Sagecal} is still only mildly enforced at the DI-calibration step (i.e.\ the regularization strength is kept low). The reason is that at this first step in the calibration process, band-pass and cable-reflection structure in the frequency direction are still present in the data and need to be corrected. Because the signal-to-noise of the sky model is very high, spectrally-correlated calibration errors may still be introduced. It has been demonstrated that chromatic DI-calibration errors due to an imperfect sky-model can be transfered from longer to shorter baselines~\citep{Ewall17,Barry16,Patil16}. These spectrally-correlated gains, when applied to the data, can then introduce spectral fluctuations well above the foreground wedge horizon and could be an origin of our observed excess. The 1416 brightest sources in our sky model account for about 99\% total sky model power, suggesting the leakage is most probably relatively small. However its impact on the power-spectra is difficult to evaluate without proper simulations, because of the spectrally-correlated errors sky-residuals  introduce~\citep{Datta10,Ewall17,Barry16}. We plan to perform such simulations in future work, although the impact of sky-incompleteness has theoretically already been analysed, in a LOFAR-like setup, by~\citet{Sardarabadi19}, as discussed earlier.

  \item[RFI ---] Low-level RFI may still pass undetected by \texttt{AOflagger}~\citep{Wilensky19}. It is currently applied on $\approx$12.2\,kHz frequency which is not optimal for detecting low-level narrow-band RFI. The additional flagging operation that is applied to the gridded visibilities cube may also miss such RFI. Faint broadband RFI could also introduce frequency structure at high $k_{\parallel}$ and is usually difficult to detect and flag. However, it would be difficult to explain the LST dependency of the night-to-night correlation.

   \lvek{Conclusion? Do we believe this is the dominant cause at this point?} \fms{Difficult to conclude firmly I would say.}
\lvekhidden{Where? Reference or figure?} 
  
  \item[Intrinsic spectral structure in the data and instrument ---] Our calibration strategy assumes that direction-dependent effects are spectrally smooth and relatively stable in time (we use a time solution interval of $2.5 - 20$ min). Some effects, such as ionospheric scintillation noise, which have decorrelation times of the order of seconds~\citep{Vedantham16}, are not solved and can leave frequency correlated noise. Scintillation noise due to bright sources such as Cas~A and Cygnus~A could also scatter power at high $k_{\parallel}$, above the `foregrounds wedge'~\citep[e.g.][]{Gehlot18}. Spectral structure in the signal chain of the instrument~\citep{Beardsley16,kern19a} is another source of spectrally correlated errors. It is however quite stable between nights and thus calibratable. \lvek{What about spectral power that is real? For example ionospheric scintillation noise to the bright source such as Cas A and Cygn A as seen in LBA data of Bharat [reference?], hence not errors via gain by power directly scattered in to the data which can therefore not be solved by smoothing the gains. Secondly, are we sure there is not spectral structure in the instrument that is different between different receivers and hence can not be solved either? The latter is not likely since it would be stable between nights and probably be calibratable, but the former might still be an option and would not strongly correlate between nights. We need to discuss both there} \fms{Any other points here ?}

\end{description}

\lvekhidden{I moved the future improvements to the conclusion section. I'm not sure it should be here.}

\noindent Most likely the excess power is not due to just one of the above
causes, but to a combination.

\subsection{Future data-processing enhancements}
\label{sec:next_steps}

Most of the causes of excess power that we discussed in the previous section
could be mitigated by improving RFI mitigation, the instrumental and ionosphere calibration scheme, our sky
model and the GPR covariance model:

\begin{description}
  \setlength\itemsep{1em}

  \item[Improving the low level RFI flagging ---] Currently about 5\% of the $uv$-cells and several sub-bands are flagged after gridding. If this low-level RFI could be flagged on higher resolution datasets, this could improve our sensitivity and reduce their impact in the EoR window. Combining the time-differenced visibilities amplitude of all baselines, a technique recently introduced in~\cite{Wilensky19}, will be used to identify faint RFI below the single baseline thermal noise. Ground-plan sources of broadband RFI will also be investigated and suppressed using near-field imaging~\citep[e.g.][]{Paciga11}.

  \item[Enforcing spectrally-smooth solutions at the DI steps ---] This is not done right now and could still lead to small chromatic gain calibration errors. In this process, we will have to separately fit slowly time varying band-pass effects, such as cable reflections, which would not be modeled by the Bernstein polynomial prior. A second DI-calibration step with a long solution time and low regularization (i.e. bandpass calibration) would be able to solve them with limited extra noise~\citep[e.g.][]{Barry19,Li19}. We will also investigate directly using the Bernstein polynomial prior as gain solutions at the DI and DD steps which could reduce chromatic gain errors and the over-fitting effect even further. This will also mitigate the impact of having an incomplete sky model~\citep{Barry16,Ewall17}.

  \item[Improving the GPR covariance model ---] The GPR method requires a covariance model that is a good statistical description of the data to be effective. Covariance kernels that would better describe the foreground wedge and the 21-cm signal would improve this model. This requires building a physically motivated spectral and spatial covariance model for each source of mode-mixing contaminant (calibration errors, ionosphere, instrument chromaticity, ...) and building a 21-cm signal covariance model, directly parameterized with EoR physical parameters.

 \item[Optimizing \texttt{Sagecal} calibration settings ---] We will also revise the solution times of the DD-calibration, the order of Bernstein polynomial prior and the maximum baselines used in the calibration. Decoupling the phase and amplitude solution time intervals could also further reduce calibration errors.

 \item[Improving the NCP sky model ---] Finally, a complete review of our current sky model will be carried out, investigating as well the inclusion of diffuse Stokes I, Q and U emission as observed using the AARTFAAC\footnote{Amsterdam-ASTRON Radio Transients Facility and Analysis Center} HBA system~\citep{Prasad16,Gehlot19t}.

\end{description}




\section{Conclusions}
\label{sec:summary}

The LOFAR-EoR KSP's primary objective is to detect the 21-cm signal from the Epoch of Reionization in the redshift range $z \approx 7 - 11$. We expect that a total of at least 1000 hours of observation with the LOFAR-HBA system will be necessary for a detection of the signal predicated by a wide range of theoretical models~\citep{Mertens18}. Whereas in~\cite{Patil17} we presented a first upper limit from one night of data (13\,h), in this work we processed twelve nights of data, combining the best ten nights (141\,hours). Compared to~\cite{Patil17}, we have introduced significant enhancements in the direction-dependent calibration of the data, replaced the foregrounds mitigation strategy and improved the power spectra extraction, leading to significantly deeper limits on the 21-cm signal even when using the same data. Our main results are the following:

\begin{enumerate}[label=(\arabic*),align=left]
  \setlength{\itemsep}{0.4em}

  \item The excess power, due to gain over-fitting (see~\citealt{Patil16} for an extensive discussion), that appears on short baselines when a baseline cut is introduced between the imaging and calibration steps\footnote{I.e. removing baselines $<$250$\,\lambda$ during calibration and only imaging 50-250\,$\lambda$ baselines during the 21-cm signal analysis phase.}, has been considerably reduced by increasing (via regularisation) the spectral smoothness of the gain solutions in the DD-calibration step.  The ratio of the variance between adjacent sub-band differences and thermal noise power (based on visibility differences on a 10\,s time scale) is reduced to a factor of $\approx 1.8$ from a factor $\approx 10$ in the procedure used in~\cite{Patil17}. In addition, we introduced Gaussian Process Regression (GPR;~\cite{Mertens18}) to remove the residual foreground emission after sky-model subtraction in the DD-calibration step. We find GPR to be more suitable compared to the Generalized Morphological Component Analysis (GMCA) method~\citep{Chapman13} in the implementation used by~\citet{Patil17}.

  \item We analysed data from twelve nights of observation obtained during LOFAR Cycles 0, 1 and 2. The data quality was found to be similar from night to night, except for two nights that were discarded from the final analysis. In all data sets, spectrally-uncorrelated (white power spectrum) noise on baselines $<100\,\lambda$ is larger than expected for thermal noise (by up to a factor 2 to 3). It is seen in both Stokes I and Stokes V, and does not appear to be related to the calibration, sky foregrounds or polarization leakage, in any clear way. Low-level RFI, below the flagging threshold, could be a possible cause of this particular white excess noise on very short baselines. Further examination and mitigation of the excess noise is planned. 

  \lvekhidden{Which ones?}
  \lvekhidden{Quote by how much?}

  \item After foreground removal using both DD-calibration and GPR, the Stokes-I residual power spectrum is characterized by a spectrally-correlated excess which is included in the overall GPR covariance model as a Matern kernel. It has a coherence scale $l_{\mathrm{ex}} \approx 0.25 - 0.45$ MHz, depending on the night. This excess is partially correlated between nights, especially in the foreground wedge region but also outside it. Larger correlations are also found between observations that started at similar LST times. The latter finding and the relatively rapid spectral de-correlation, together suggest that the residuals may originate from un-modelled or incorrectly modelled sky emission far from the phase center.

  \lvekhidden{Prior or range after optimizing? And this range is the spread between nights?}

  \item After combining the best 10 out of 12 analysed nights of data (141\'h of data), the residual Stokes-I power decreased by a factor of $\approx$4 in the foreground wedge region, and by a factor of $5 - 6$ outside of the wedge. The residuals are dominated by the same spectrally correlated excess noise found in all individual nights.

  \item Based on the 141\,h data set, we find an improved $2-\sigma$ upper limit on the 21-cm signal power spectrum at $z \approx 9.1$ of $\Delta^2_{21} <
  (72.86)^2\, \mathrm{mK^2}$ at $k = 0.075 \mathrm{h\,cMpc^{-1}}$ (the lowest $k$-mode) and
  $\Delta^2_{21} < (106.65)^2\, \mathrm{mK^2}$ at $k = 0.1 \mathrm{h\,cMpc^{-1}}$ (the reference $k$-mode), with a  $\mathrm{d}k / k \approx 0.3$. The latter is an improvement by a factor $\approx 8$ in power compared to the previous upper limit reported in~\cite{Patil17}. 


  \lvekhidden{Indicate the $\delta k/k$ here for clarity. If this is $\sim 0.3$ these limits might still go down when using $\delta k/k \sim 1$. We could consider quoting numbers for the latter bin-size as well. }

  \item We have examined a range of possible origins for the excess power, including residual foregrounds emission from sources away from the phase center, polarization leakage of Stokes Q and U emission to Stokes I, chromatic DI/DD-calibration errors and low level RFI. No clear cause has yet been identified, but further improvements of our processing procedures are currently under way to reduce its level by (i) improving low level RFI flagging, (ii) enforcing spectrally-smooth solutions during DI-calibration, (iii) further optimizing \texttt{Sagecal} calibration settings (regularization prior, number of ADMM iterations, applying the Bernstein polynomial prior itself instead of the regularised gain solutions) and (iv) using more physically motivated GPR covariance models that are not only defined in the frequency direction, but also in time and baseline, to better separate the various contributions to the power spectrum and 21-cm signal limits.

  \item Based on current estimates of the thermal noise in the analysed data sets, which we believe to be accurate, and assuming that the excess power can be mitigated, one can reach a $2-\sigma$ sensitivity limit of $\approx (14)^2\, \mathrm{mK^2}$ at $k = 0.1 
  \mathrm{h\,cMpc^{-1}}$ from the same 10 nights of data, and a very deep $\approx (4)^2\, \mathrm{mK^2}$ sensitivity limit, when combining about 100 nights of data, which is in the range where current 21-cm EoR models predict the power to be.

  \lvekhidden{State that this in within the regime where models predict the power to be. Also here we are conservative since in general people assume $\delta k/k \approx 1$. So we can go another $\sqrt{3}$ lower if we assume a larger bins size. Also we have multiple redshifts that can be combined when comparing to models. Maybe comment on this.}

\end{enumerate}

Although the cause of excess noise has still not been fully solved, the results presented in this paper are a significant step forward compared to those by~\citet{Patil17}. Several issues that were identified in that work have now largely been mitigated, and a number of major improvements in our data processing procedure have been achieved. In the present analysis, possible sources of the  excess power have been unveiled and solutions to mitigate them are currently investigated. 

\subsection{Implication of the upper limit on the EoR}
\label{sec:implication}

The implications of the improved 21-cm signal power spectrum upper limit on the Epoch of Heating (EoH) and Epoch of Reionization (EoR) are analyzed in detail in an accompanying paper by~\cite{Ghara20} using the reionization simulation code GRIZZLY~\citep{Ghara15a,Ghara18} and a Bayesian inference framework to constrain the parameters of the IGM. They study two sets of extreme scenarios that can be constrained by this upper limit: (i) For an IGM with a uniform spin temperature, they find that the models which can be ruled out have a combination of a very cold IGM (spin temperature $< 3\,\mathrm{K}$) and a high UV photon emission rate~\citep{Ghara20}. (ii) In the case of a non-uniform IGM spin temperature, they find that the current upper limit is likely to rule out models with large emission regions which do not cover more than a third of an otherwise unheated IGM~\citep{Ghara20}.



\section*{Acknowledgements}
FGM and LVEK  would like to acknowledge support from a SKA-NL Roadmap grant from the Dutch ministry of OCW. 
SZ acknowledges support from the Israeli Science Foundation (grant no. 255/18). 
ITI was supported by the Science and Technology Facilities Council [grants ST/I000976/1, ST/F002858/1, and ST/P000525/1]; and The Southeast Physics Network (SEPNet). 
GM is thankful for support by Swedish Research Council grant 2016-03581. 
VJ acknowledges support by the Croatian Science Foundation for a project IP-2018-01-2889 (LowFreqCRO).
EC acknowledges support from the Royal Society via. the Dorothy Hodgkin Fellowship.




\bibliographystyle{mnras}
\bibliography{references}

\clearpage

\appendix

\section{Signal injection tests and simulations}
\label{sec:inj_simu_test}

GPR foreground mitigation may alter the 21-cm signal and assessing its efficiency and robustness is therefore crucial. In~\cite{Mertens18} we have carried out numerous tests against a large range of foregrounds simulations. Here we present tests which are more specifically connected to the frequency correlations observed in the LOFAR data.

\lvekhidden{Cite your paper here and say that some tests were already carried out there and what the difference is with the tests here.}

\begin{figure}
    \includegraphics{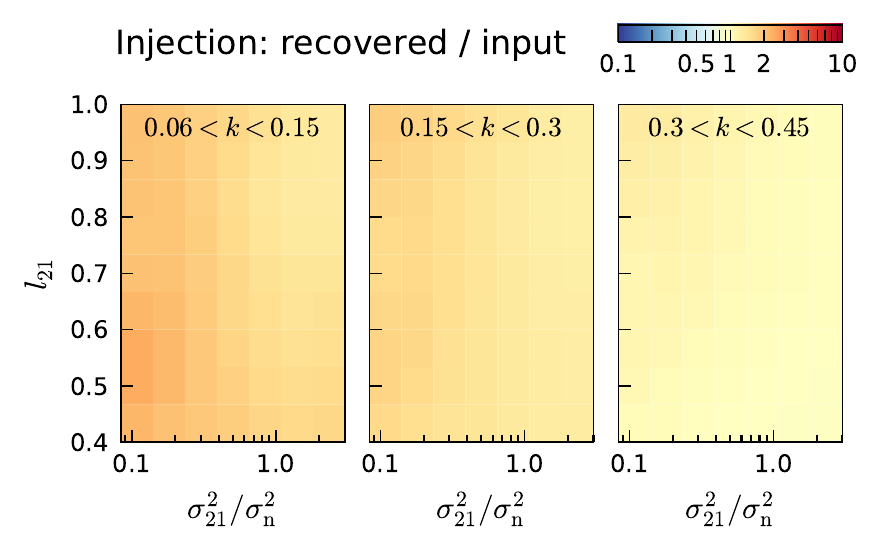}
    \caption{\label{fig:gpr_injection_test}Result of the injection test
    for a wide range of coherence scale ($l_{21}$) and S/N ($\sigma^2_{21} /
        \sigma^2_{\mathrm{n}}$) of the 21-cm like injected simulated signal. We plot the ratio
    of the recovered over injected signal spherically-averaged power
    spectra for three $k$-bins.}
\end{figure}

\noindent {\it Signal injection in real data -- } One way to do this is by injecting artificial 21-cm  signals into real data and comparing the GPR results to those without the additional 21-cm-like signal. Denoting the matrix $\rm P$ as the GPR foreground-mitigation (projection) operator, applied to the data ($\mathbf v$), we obtain the recovered signal by taking the difference between the two processed data sets:
\begin{equation}
\mathbf{v}_{\mathrm{rec}} = {\rm P'}(\mathbf{v}_{\mathrm{data}} + \mathbf{v}_{\mathrm{inj}}) -{\rm P}\, \mathbf{v}_{\mathrm{data}}.
\end{equation}
The prime denotes here that the GP model parameters were re-optimized for the data set with the injected signal.
\lvekhidden{It looked before that the FG mitigation functions where the same. I presume you redo the full MCMC analysis and hence the operators have different parameters. I denoted this by a prime on one of the operators. Please make clear that you reoptimize the kernels and also that both the data w/o 21-cm signal and the data plus 21-cm have the same covariance kernels, so the 21-cm kernel is also used when no 21-cm signal is injected.} 
The 21-cm signals are approximated by an exponential covariance function~\citep{Mertens18}. Figure~\ref{fig:gpr_injection_test} presents the ratio of the spherically-averaged power spectra from the recovered over the injected 21-cm signals for a wide range of coherence scales and variances of the injected signal. For each combination of these variables, we perform 10 simulations and the result is averaged. A ratio of 1.0 indicates no bias and a ratio $< 1$ indicates signal loss. We note that all bias values, when found, are strictly confined to the regime $>1$ and are limited to larger coherence scales and smaller signal-to-noise ratios. 

\lvekhidden{Cite your paper where you show this is a decent approximation and state the latter.}

\begin{figure}
    \includegraphics{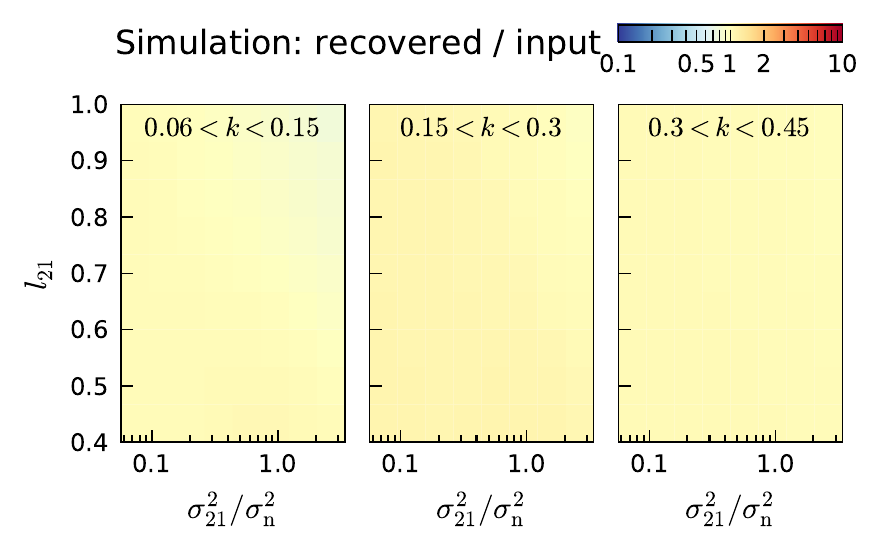}
    \caption{\label{fig:gpr_simulation_test}Result of the simulation test for a
    wide range of coherence scale $l_{21}$ and S/N  ($\sigma^2_{21} /
        \sigma^2_{\mathrm{n}}$) of the simulated 21-cm like signal. We plot the
        ratio of the recovered over injected signal spherically-averaged power
        spectra for three $k$-bins. In this case, the recovered and input
        include the `excess' signal.}
\end{figure}

\noindent {\it Signal injection in simulated data -- } We also perform data simulations that reproduce the spectral correlations found in the full data set, using its optimal GPR covariance model parameters. 
For these simulations, our input `signal' is the `21-cm signal' and `excess'. GPR is
applied to these data sets using a similar setup as for the injection tests, and
we compute the ratio of the recovered over input power spectra. Our results
(Figure~\ref{fig:gpr_simulation_test}) show a ratio $\approx 1$ for all the tested
coherence scales and S/N of the 21-cm signal.

\section{Confidence interval on the GP model hyper-parameters}
\label{sec:mcmc}

\begin{figure*}
    \includegraphics{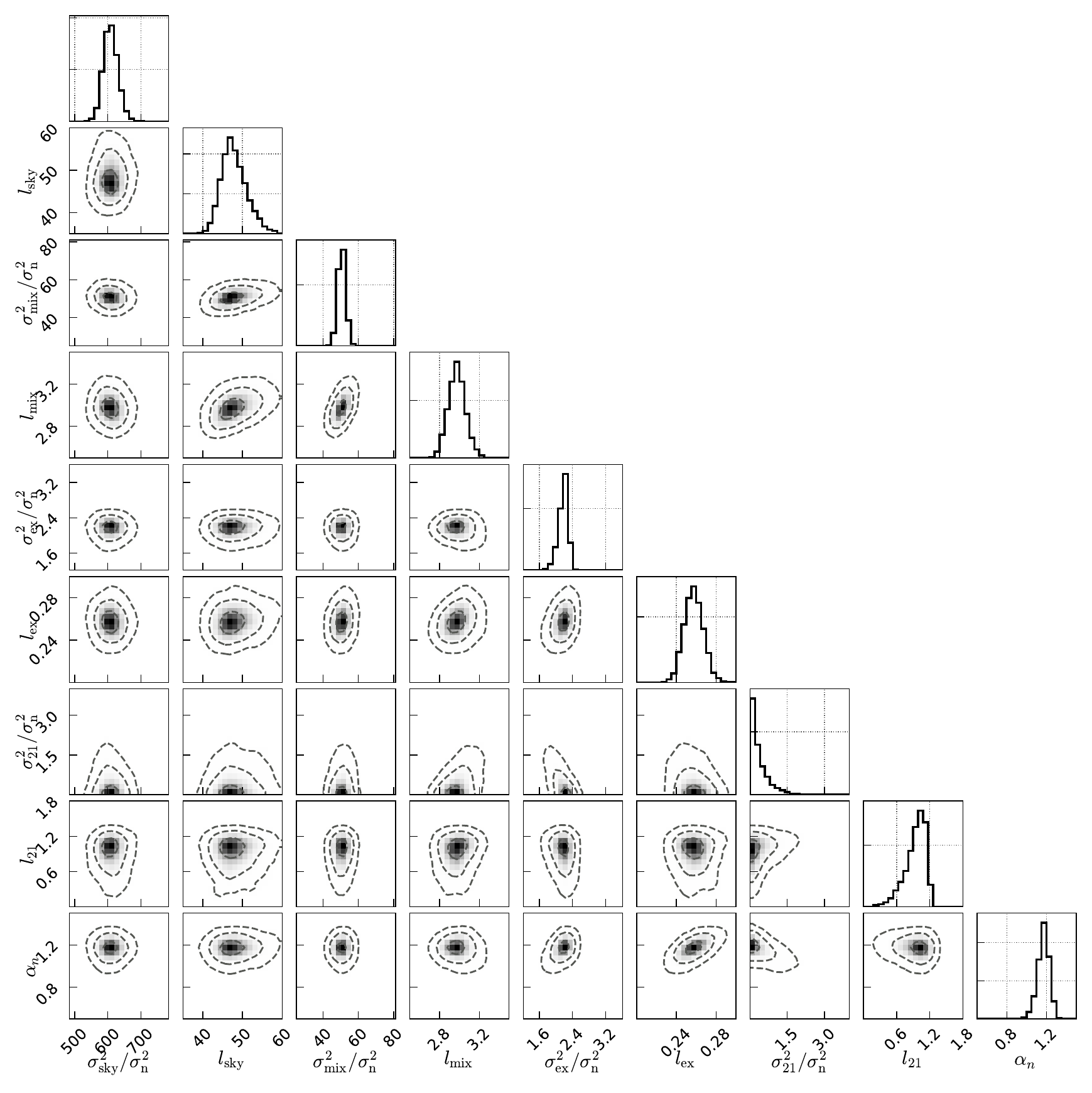}
    \caption{\label{fig:gpr_mcmc_9params}Posterior probability distributions
    of the GP model hyper-parameters for the 10 nights dataset. The covariance
    model has 9 parameters: 2 for each of the \textit{sky}, \textit{mix},
    \textit{21} and \textit{ex} (excess) components, plus the scaling factor
    $\alpha_n$. The black dashed contours show the 68\%, 95\% and 99.7\% confidence
    interval.}
\end{figure*}

A Monte Carlo Markov Chain (MCMC) can be used to fully sample the posterior distribution of the GP model's hyper-parameters. This allows us to validate the optimal values obtained by optimization algorithm, and to estimate their confidence intervals. We apply the MCMC method\footnote{This procedure uses the~\texttt{emcee} python package (http://dfm.io/emcee/current/)~\citep{Foreman-Mackey13}.} described in Section 4.2.2 of~\cite{Mertens18} on the 10 nights data set. Figure~\ref{fig:gpr_mcmc_9params} shows the resulting posterior probability distribution of the GP model hyper-parameters. The parameter estimates and confidence intervals are summarized in Table~\ref{tab:gp_model_fit}, along with their input values and associated priors. The correlation between the different parameters of the model is overall very small. All parameters are also well constrained, except the variance of the `21-cm signal' component, which is consistent with zero.

\bsp    
\label{lastpage}

\end{document}